\documentclass[conference]{IEEEtran}
\IEEEoverridecommandlockouts
\usepackage{cite}
\usepackage{amsmath,amssymb,amsfonts}
\usepackage{graphicx}
\usepackage{textcomp}
\usepackage{xcolor}
\usepackage{comment}
\usepackage{tabularx}
\usepackage[T1]{fontenc}
\usepackage{subcaption}
\usepackage{float}
\usepackage{amsmath}
\usepackage{microtype}
\usepackage[linesnumbered,ruled,vlined]{algorithm2e}
\usepackage{tikz}
\usepackage{hyperref}
\usepackage{placeins}
\usetikzlibrary{positioning}
\SetKwBlock{Loop}{Loop}{EndLoop}

\def\BibTeX{{\rm B\kern-.05em{\sc i\kern-.025em b}\kern-.08em
    T\kern-.1667em\lower.7ex\hbox{E}\kern-.125emX}}

\makeatletter
\newcommand{\removelatexerror}{\let\@latex@error\@gobble}
\makeatother

\begin{document}

\title{Emergent Peer-to-Peer Multi-Hub Topology\thanks{The work presented in this document has received funding from the EU
Horizon Europe research and innovation Programme under Grant Agreement No. 101070118.}}

\author{\IEEEauthorblockN{Mohamed Amine LEGHERABA}
\IEEEauthorblockA{\textit{LIP6-NPA} \\
\textit{Sorbonne University}\\
Paris, France \\
mohamed.legheraba@lip6.fr}
\and
\IEEEauthorblockN{Maria POTOP-BUTUCARU}
\IEEEauthorblockA{\textit{LIP6-NPA} \\
\textit{Sorbonne University}\\
Paris, France \\
maria.potop-butucaru@lip6.fr}
\and
\IEEEauthorblockN{Sébastien TIXEUIL}
\IEEEauthorblockA{\textit{LIP6-NPA} \\
\textit{Sorbonne University}\\
\textit{IUF}\\
Paris, France \\
sebastien.tixeuil@lip6.fr}
\and
\IEEEauthorblockN{Serge FDIDA}
\IEEEauthorblockA{\textit{LIP6-NPA} \\
\textit{Sorbonne University}\\
Paris, France \\
serge.fdida@lip6.fr}}

\maketitle
\thispagestyle{plain}

\begin{abstract}

In this paper we propose and evaluate an innovative algorithm that enables the creation of Peer-to-Peer network overlays characterized by emergent multi-hubs. This approach generates overlays that balance between the randomness of a graph and the structure of a star network, resulting in networks that not only feature prominent hubs but also exhibit strong resilience to failures. By leveraging principles of preferential attachment and random attachment, our method allows hubs to form spontaneously, offering a decentralized and fault-tolerant solution ideal for applications requiring both low network diameter and high robustness. The protocol is entirely decentralized, operates asynchronously, and depends exclusively on local information. Nodes organically evolve into hubs and remain indistinguishable from other nodes (except in terms of the number of incoming links). The quantity of hubs that emerge can be predetermined by the application as a network parameter.
\end{abstract}


\section{Introduction}
\label{sec:1}
    The growing usage of decentralized systems such as blockchain~\cite{bitcoin2008bitcoin} and federated learning~\cite{mcmahan2017communication} in recent years has sparked considerable interest in peer-to-peer (P2P) communication protocols. While existing P2P protocols have demonstrated significant utility across various applications, emerging demands for enhanced performance, scalability, and robustness necessitate the development of innovative solutions.

Peer-to-peer (P2P) protocols have undergone extensive research and development to facilitate efficient decentralized communication among networked devices. Foundational P2P protocols like Napster, Gnutella~\cite{frankel2003gnutella}, and BitTorrent paved the way for distributed file sharing and content distribution across the Internet. Typically, P2P overlay networks are categorized as either structured (e.g. CAN~\cite{ratnasamy2001scalable}, Chord~\cite{stoica2003chord}, or Kademlia~\cite{maymounkov2002kademlia}) or unstructured (e.g. Gnutella~\cite{frankel2003gnutella}). More comprehensive details about peer-to-peer overlays can be found in recent surveys~\cite{malatras2015state, naik2020next}. 

Structured overlays come with a maintenance cost~\cite{malatras2015state}, and are more susceptible to Byzantine attacks (that is, attacks performed by the peers themselves)~\cite{naik2020next} and churn~\cite{malatras2015state} (that is, the unexpected departure and arrival process of the peers). 
Unstructured networks exhibit advantages in resilience to node failures and adaptability to shifting network conditions~\cite{jelasity2007gossip}, rendering them well-suited for dynamic and heterogeneous environments when compared to their structured counterparts. Their shortcomings are that the quality of services built on top of the network is difficult to assess. 

\subsection{Peer sampling.}
Peers within an unstructured overlay maintain a dynamic set of neighbors, often discovered through mechanisms like peer sampling~\cite{jelasity2007gossip}, which enables nodes to gather and exchange information about other nodes in the network, and thus dictates the network topology.
Existing peer sampling algorithms in the literature yield two types of topologies (random and power-law) that demonstrate favorable networking characteristics. Random graphs are built from gossip peer sampling algorithms and are known to be resilient to churn~\cite{jelasity2007gossip}. 
Power-law (or scale-free) networks are built from algorithms that use the concept of preferential attachment and are known to have ultra-small diameter~\cite{cohen2003scale}, which helps scalability. 
However, when considering the specific use case of federated learning, certain limitations emerge: \emph{(i)} gossip learning, based on gossip peer sampling, exhibits a slower convergence rate compared to centralized federated learning methodologies~\cite{hegedHus2021decentralized}, and \emph{(ii)} while power-law topologies theoretically offer improved convergence efficiency, prior research has predominantly focused on constructing networks adhering strictly to power-law distributions~\cite{xie2008scale, bulut2013constructing}, implementing algorithms to restrict the proliferation of hubs~\cite{guclu2008limited, eum2009self} (that is, peers that are extremely well connected), or leveraging other metrics to construct node connections, like the distance in terms of Internet hops~\cite{sasabe2006llr} or an initial attractiveness~\cite{park2018distributed}.

Yet, for federated learning, the presence of hubs is advantageous, as these hubs facilitate rapid relay of machine learning models across the network, accelerating convergence rates. Nonetheless, conventional approaches relying on predefined hubs (e.g., super-peer-based topologies) are susceptible to attacks targeting static and well-defined hub nodes~\cite{montresor2004robust}.

Hence, there exists a pressing need for a protocol that fosters the organic emergence of hubs within networks. The service outlined in this article is designed precisely for this purpose, allowing selected nodes to naturally ascend to hub status through a process we term "hub sampling". By enabling nodes to organically assume the role of hubs, our protocol aims to strike a balance between leveraging the efficiency of hub-based networks for applications like federated learning, while mitigating vulnerabilities associated with static hub designations.

\subsection{Our contribution.} Our primary goal is to develop a protocol that autonomously promotes nodes to act as hubs within unstructured peer-to-peer networks. To achieve this goal, we hybridize two fundamental concepts: \emph{preferential} attachment, and \emph{random attachment}.
By integrating these two concepts, our protocol promotes a balanced network structure, where hubs emerge organically based on connectivity patterns and yet adapt to dynamic network changes. 
This approach not only fosters robustness against failures and disruptions but also maintains a low network diameter, facilitating efficient communication and information propagation. The parameter \emph{h}, representing the desired number of hubs, allows for flexibility and control over the network's topology, enabling tailored configurations to suit specific application requirements and network environments.
The rationale behind this initiative is rooted in the benefits of having hub nodes, particularly in applications such as federated learning, where efficient information dissemination is crucial. The existence of hubs facilitates faster network-wide communication compared to overlay networks structured in a random graph topology.

The structure of this article is organized as follows: Section~\ref{sec:2} presents the hub sampling service altogether with its properties, its programming interface (API), and its implementation, the \emph{Elevator algorithm}. Section~\ref{sec:3} presents a theoretical analysis of the properties of the algorithm. Section~\ref{sec:4} presents extensive simulations of Elevator, compared against three classical algorithms from the literature~\cite{jelasity2007gossip,stavrou2004lightweight,wouhaybi2004phenix}.

\section{Hub sampling service}
\label{sec:2}

The key desired properties we expect from our protocol are \emph{connectivity} (the overlay remains connected), \emph{low-diameter} (for efficient communication), \emph{convergence} (properties are obtained in an autonomous manner), \emph{stability} (structural overlay properties are maintained throughout execution), and \emph{robustness} (resilience to churn and targeted attacks). They will serve as metrics during simulation experiments to ascertain the efficacy of our algorithm.

\subsection{Service API}
The API of the hub sampling service mirrors that of classical peer sampling service~\cite{jelasity2007gossip}, comprising two key methods:
    \emph{(i)} \emph{init()} that initializes the service on a given node, \emph{i.e.}, initializes the list of outgoing connections of a node (Indeed, we assume that a given node starts connected to a random subset of nodes in the network, the actual initialization procedure being implementation-dependent), and
    \emph{(ii)} \emph{getPeer()} that returns a random peer address from the node list of peers.

The focus of this work is to present an implementation of the \emph{getPeer()} method, Elevator, as a gossip-based algorithm, and to study the performance of its implementation. 
In addition to these two methods, we add a third method to the API called \emph{getHub()} that returns a random hub. The \emph{getHub()} method can be easily derived from \emph{getPeer()} by filtering the output of \emph{getPeer()} to only select the $h$ nodes acting as hubs in the network. This method can be useful for applications that only need to contact a hub.

\subsection{Preliminaries}
In the context of our study, we consider an overlay network of interconnected nodes modeled as a directed graph. Communication within this network is bidirectional, corresponding to an underlying undirected graph that represents the physical network. Each node in this network possesses a unique address, akin to an IP address in the context of the Internet, serving as an abstract identifier of its identity. Nodes maintain a local list called \emph{cache}, which contains addresses of other nodes, and represents their partial knowledge of the network's node set. The maximum size of this cache, denoted by parameter \emph{c}, is uniform across all nodes. The cache is pivotal for peer sampling, as it serves as the basis for neighbor selection and information exchange. At the network's inception, nodes are initially connected to a random subset of nodes, forming what is known as a random \emph{k}-out graph. Subsequently, new nodes joining the network also establish connections with a random subset of existing nodes, a process that populates their cache and integrates them into the network. Given the decentralized nature of the network, peer sampling algorithms are designed to operate asynchronously, as it is the case for Elevator, and all algorithms presented in this paper, but to help the evaluation of protocols during simulations, we can refer to the idea of \emph{cycles} of the protocol. During each cycle, every node initiates one execution of the peer sampling protocol, potentially updating its cache based on interactions with neighboring nodes. By leveraging cycles, we can analyze the convergence, performance, and robustness of peer sampling protocols under varying conditions and scenarios within the decentralized network environment.

\subsection{Elevator core concepts}
To achieve both robustness and a low network diameter, we integrate two fundamental concepts: preferential attachment and random attachment, each serving distinct yet complementary roles in shaping the network topology.

\textbf{Preferential Attachment.} Drawing from the concept pioneered by Barabási and Albert~\cite{barabasi2002evolution}, preferential attachment dictates that new connections in the network are established preferentially with nodes possessing a higher number of existing connections. In our adaptation, we modify this concept to elevate certain nodes to the status of hubs without requiring the network to continuously grow. Instead of new nodes joining and preferentially connecting to highly connected nodes, each existing node leverages information from its neighbors to identify and connect to the most frequently connected nodes (up to a predefined number \emph{h}). This mechanism enables the organic emergence of hubs within the network, with selected nodes naturally assuming central roles based on their connectivity without any explicit distinction other than their number of incoming links.

\textbf{Random Attachment.} Inspired by gossip-based peer sampling algorithms~\cite{jelasity2007gossip, stavrou2004lightweight}, random attachment ensures that nodes maintain connections with a representative and diverse subset of the network. This strategy promotes network robustness by preventing excessive clustering and dependency on specific nodes (hubs). When existing hubs disappear (e.g., due to failures or departure), other nodes within the network are opportunistically elevated to hub status, ensuring continuity and adaptability of the network topology over time.

Our target is to obtain a topology of the network that has the following properties:
    \emph{(i)} There are \emph{h} defined hubs, with \emph{h} a parameter defined before the start of the network and common to all nodes,
    \emph{(ii)} ignoring hubs, the distribution of the remaining connections is random, and
    \emph{(iii)} each node has \emph{c} connections, consisting of \emph{h} connections to hubs and \emph{c-h} connections to random nodes.

Through simulation evaluation, we demonstrate in the sequel the effectiveness and advantages of our protocol with respect to state-of-the-art algorithms.

\subsection{Elevator detailed description}
The algorithm uses the following parameters and data structures:
\begin{itemize}
    \item \emph{Parameter $c$}: The maximum number of outgoing connections (its default value for all nodes is 20). 
    \item \emph{Parameter $h$}: The number of preferential attachment connections (its default value for all nodes is \emph{c/2}).
    \item  \emph{Parameter $\mathit{maxsize\_buffer\_backward}$}: The maximum number of backward connections to send (its default value for all nodes is 100).
    \item \emph{Structure $\mathit{cache}$}: The list of outgoing connections. The list is implemented as an array of size \emph{c}. The list is initialized with random existing addresses (random connections to other nodes of the network).
    \item \emph{Structure $\mathit{backward\_peers}$}: The list of other nodes that have tried to connect to the node. The list is implemented as a linked list (initially empty).
\end{itemize}
Additionally, we have three temporary structures: \emph{(i)} \emph{$\mathit{frequency\_map}$} holds the frequency of occurrences for all neighbors of neighbors, implemented as a map ($\mathit{node} \rightarrow \mathit{integer}$), \emph{(ii)} \emph{preferred} holds the list of preferred nodes, implemented as a linked list, and \emph{(iii)} \emph{$\mathit{preferred\_backward}$} holds the list of backward connections of the preferred nodes, implemented as a linked list. 

\subsubsection{The proposed protocol executes the following actions at each run} Each node retrieves the neighbor's list of their neighbors (\emph{i.e.}, the neighbors at distance two). The node then builds an ordered list of the most frequent peers (the frequency map) and contacts the \emph{c} most frequent nodes (called \emph{preferred}). Each contacted node sends back to the contacting node a maximum of \emph{$\mathit{maxsize\_buffer\_backward}$} addresses from its backward list, maintained in the structure \emph{$\mathit{backward\_peers}$}, and adds the contacting node to its backward list. The cache of the contacting node is then reset as an empty array. Then the node selects the \emph{h} most frequent peers and \emph{c-h} random peers from the list of backward peers of all preferred peers to fill its cache. If the cache is not full, the node adds random peers from the frequency map to the cache until the size of the cache is \emph{c} (see Algorithm~\ref{algo:algoLabel} and Algorithm~\ref{algo:algoLabelBackground} for detailed pseudocode of the algorithm in the Appendix \ref{sec:algoElevator}).

\section{Experimental evaluation}
\label{sec:4}

We evaluate our proposal by carrying out a simulation campaign.
All simulations use the Java \emph{PeerSim} simulator~\cite{p2p09-peersim}. 
We have modified the simulator to add parallelism to accelerate computations.
With Peersim, we implemented our algorithm Elevator, and state-of-the-art PROOFS~\cite{stavrou2004lightweight} and Phenix~\cite{wouhaybi2004phenix} algorithms~\footnote{\href{https://gitlab.lip6.fr/legheraba/elevator}{https://gitlab.lip6.fr/legheraba/elevator}}.
Also, we used the implementation of Newscast provided by PeerSim. 
A detailed description of these algorithms can be found in Appendix~\ref{sec:algorithms}.
We compared the performance of Elevator with these 3 algorithms. We chose to compare our proposed algorithm to these three algorithms as they are widely used in the literature. Newscast is used for gossip learning\cite{ormandi2013gossip}, PROOFS is a foundational algorithm, as Secure Cyclon\cite{antonov2023securecyclon}, one of the latest peer sampling algorithm in the literature, is based on Cyclon\cite{voulgaris2005cyclon}, itself based on PROOFS. Phenix is interesting as it has especially been conceived to be resilient to failures and Byzantine attacks and also to construct networks that have a low diameter. 
We did not include recent algorithms~\cite{xie2008scale, bulut2013constructing, lynn2024emergent} that primarily focus on improving the power law distribution of the in-degrees~\cite{xie2008scale, bulut2013constructing, lynn2024emergent}, as they are expected to behave similarly to Phenix~\cite{wouhaybi2004phenix}.

All simulations were run with a network of size \emph{n}=1000. As the Phenix network needs a growing network to work, we started the Phenix algorithm with a network size of 20 and capped the size of the network to 1000. The simulations were run during 1000 cycles, and we repeated each simulation 100 times.
All simulations were started with a network initialized as a $k$-out random graph, with $k$=\emph{c}=20.
All simulations were run on 16 vCPU, using 64G of memory, on a cluster composed of 10 servers, described in Table~\ref{table-cluster}.

\begin{table*}[t]
\caption{Description of the cluster}
\label{table-cluster}
\begin{center}
\begin{tabular}{ |c|c|c|c| } 
 \hline
 Machine & Memory & Processors & Cores \\ 
 DELL PowerEdge XE8545 & 2 To & 2 x AMD EPYC 7543 & 128 threads @ 2.80 GHz \\ 
 DELL PowerEdge R750xa & 2 To & 2 x Intel Xeon Gold 6330 & 112 threads @ 2.00 GHz \\ 
 \hline
\end{tabular}
\end{center}
\end{table*}

We evaluated the following metrics: in-degree distribution, clustering coefficient, average shortest path length, and diameter. More details about those classical graph metrics can be found in Appendix~\ref{sec:metrics}.

\begin{figure}[H]
    \includegraphics[width=0.85\columnwidth]{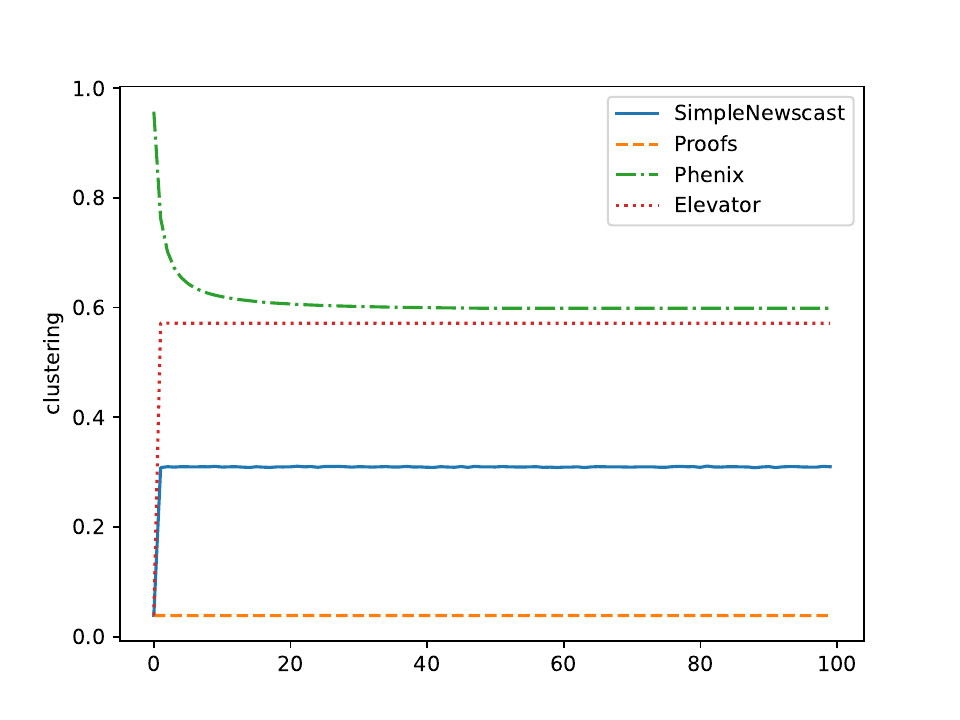}
    \caption{Clustering coefficient computed during the simulation (no failures), for each algorithm, every 10 cycles}
    \label{fig:ClustCoef}
\end{figure}

\begin{figure}[H]
    \includegraphics[width=0.85\columnwidth]{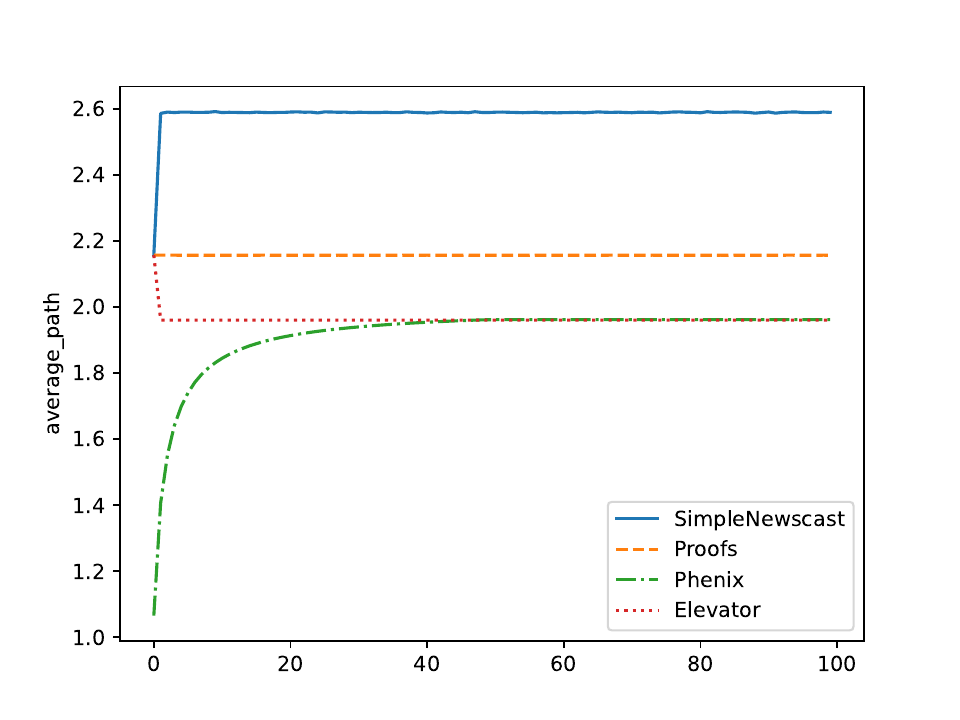}
    \caption{Average path length computed during the simulation (no failures), for each algorithm, every 10 cycles}
    \label{fig:AveragePathLength}
\end{figure}

\begin{figure}[H]
    \includegraphics[width=0.85\columnwidth]{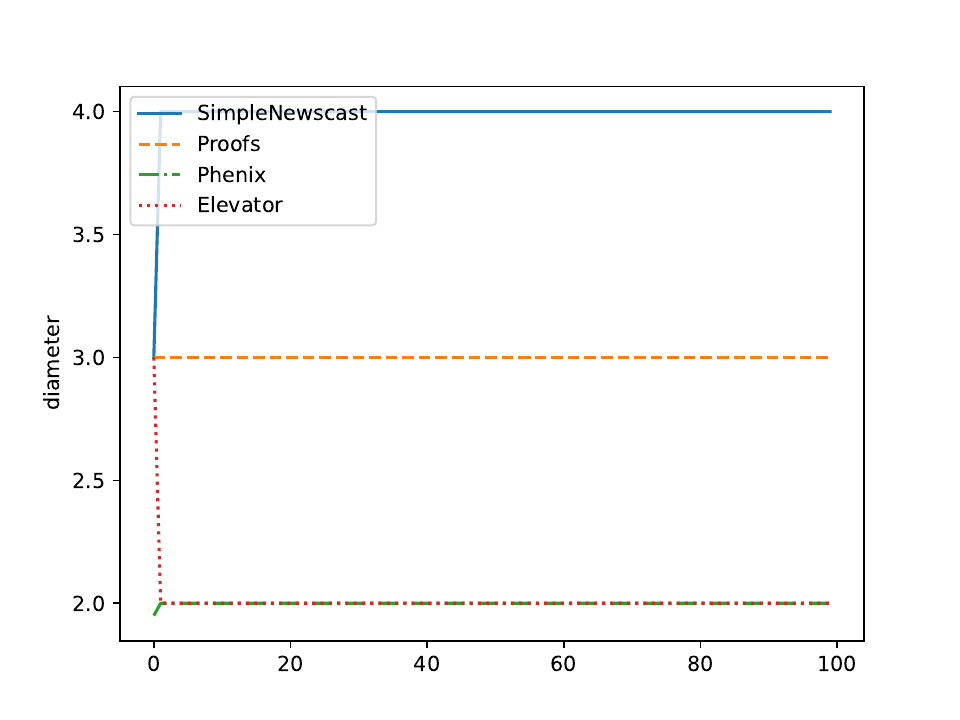}
    \caption{Diameter computed during the simulation (no failures), for each algorithm, every 10 cycles}
    \label{fig:Diameter}
\end{figure}

The degree distributions of Newscast and PROOFS exhibit patterns akin to a normal distribution. We see similar results for Elevator, except for a distinct group of 10 hubs with an in-degree of 999. By contrast, the Phenix protocol's degree distribution conforms to a power-law distribution. 
PROOFS and Newscast maintain a low clustering coefficient during all simulations, as seen in Figure~\ref{fig:ClustCoef}. On the contrary, Phenix and Elevator have both a clustering coefficient of around 0.6. For Phenix, the value is related to the power-law distribution of in-degree, and for Elevator, the value is linked to the presence of hubs, that are connected to everyone, and this automatically increases the value of the coefficient. 
As we can see in Figure~\ref{fig:AveragePathLength}, Elevator has a very low average path length, with a value below 2. This value is due to the presence of hubs in the network, that permit to have a maximum distance of 2 between any 2 nodes. Phenix has the same value. PROOFS is very close, with a value around 2.15 and Newscast is a bit below 2.6. All these values are very good and thus we need to compute the diameter to discriminate between algorithms.
In Figure~\ref{fig:Diameter}, we see that Elevator gives a network with a diameter equal to 2. Again, this value is due to the presence of hubs in the network. The Phenix algorithm yields similar results. This is better than PROOFS and Newscast, which output respectively 3 and 4 for this metric. 

We also compared the algorithms according to their resilience to crashes, churn, and attacks on hubs, as shown below. Additional results and the accompanying figures are included in the Appendix~\ref{sec:figures}. 

\subsection{Resilience to crashes}
\begin{figure}[H]
    \includegraphics[width=0.85\columnwidth]{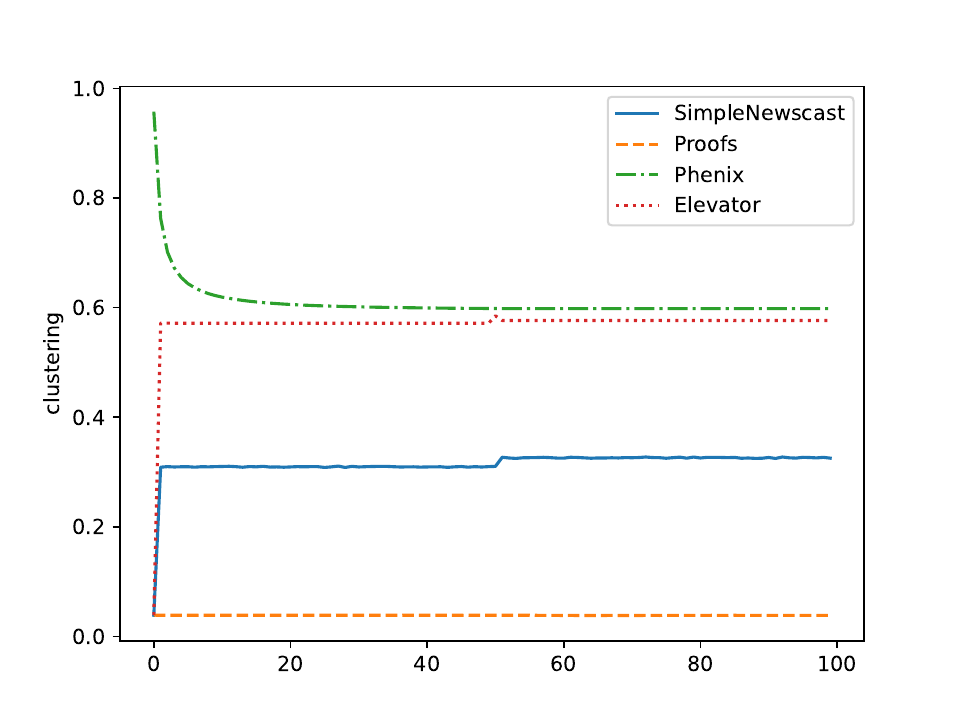}
    \caption{Clustering coefficient computed with a 50\% crash, for each algorithm, every 10 cycles}
    \label{fig:ClustCoefCrash}
\end{figure}

\begin{figure}[H]
    \includegraphics[width=0.85\columnwidth]{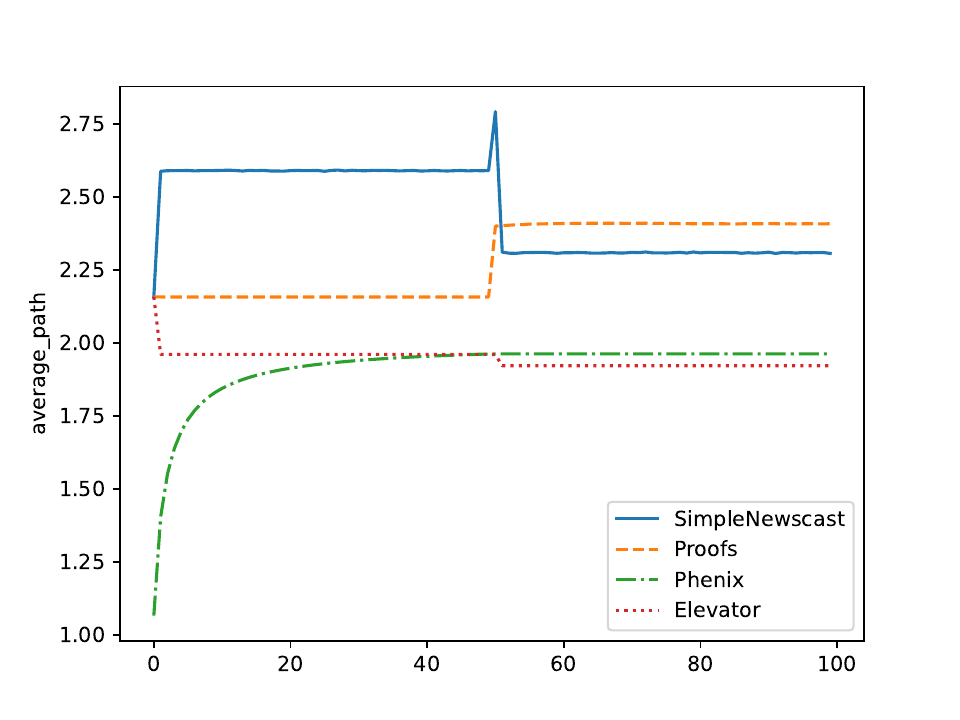}
    \caption{Average path length computed with a 50\% crash, for each algorithm, every 10 cycles}
    \label{fig:AveragePathLengthCrash}
\end{figure}

\begin{figure}[H]
    \includegraphics[width=0.85\columnwidth]{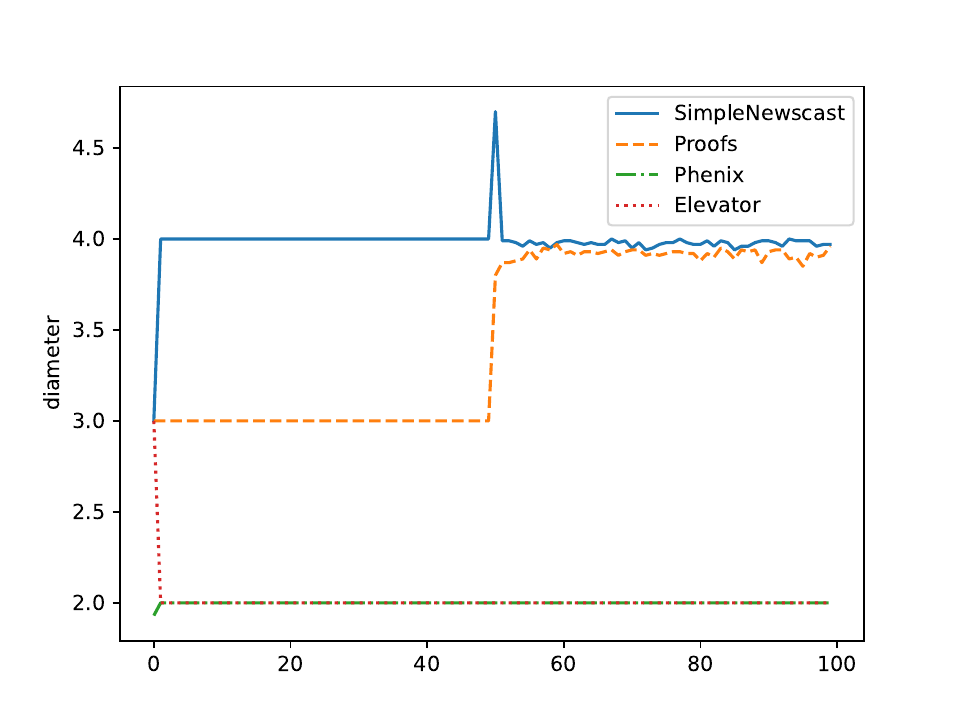}
    \caption{Diameter computed with a 50\% crash, for each algorithm, every 10 cycles}
    \label{fig:DiameterCrash}
\end{figure}

We analyze the performance of the four algorithms when the network suffers crashes. To simulate a brutal failure we disconnected 50\% of the nodes in the middle of the simulation, \emph{i.e.}, in this case, we have disconnected 500 nodes at cycle 500 (as there are 1000 nodes in total and 1000 cycles). 
The performance of Elevator is not affected, as the in-degree distribution is still the same, and we have 10 hubs with an in-degree of 499. The degree distribution is also the same for Newscast and PROOFS. For Phenix, the degree distribution remains the same, with values going to a max of 999, even if there are only 500 nodes in the network. It's because the nodes have kept in their cache the addresses of (old) nodes who are no longer in the network. In Figure~\ref{fig:ClustCoefCrash}, the clustering coefficient evolution shows that it is not affected by the crashes, as we have almost the same results as those obtained without a crash. The same observation holds for the average path length and the diameter, as we can see in Figures~\ref{fig:AveragePathLengthCrash} and \ref{fig:DiameterCrash}.

\subsection{Resilience to churn}
We now analyze the performance of the four algorithms when the network is subject to churn. To simulate churn, we disconnected 10\% of the nodes at each cycle and replaced them with the same amount of new nodes, each connected to 20 nodes uniformly at random. The churn occurs during 500 cycles, between cycle n°250 and cycle n°750. As the Phenix algorithm needs a growing network to work, the way we implement churn differs. Following previous work~\cite{wouhaybi2004phenix}, in the case of Phenix, we implement churn having the number of removed nodes less than the number of added nodes at each cycle, assuming nodes are removed following a normal distribution $\mathcal{N}(0,1)$, for all cycles of the simulation.

\begin{figure}[H]
    \includegraphics[width=0.85\columnwidth]{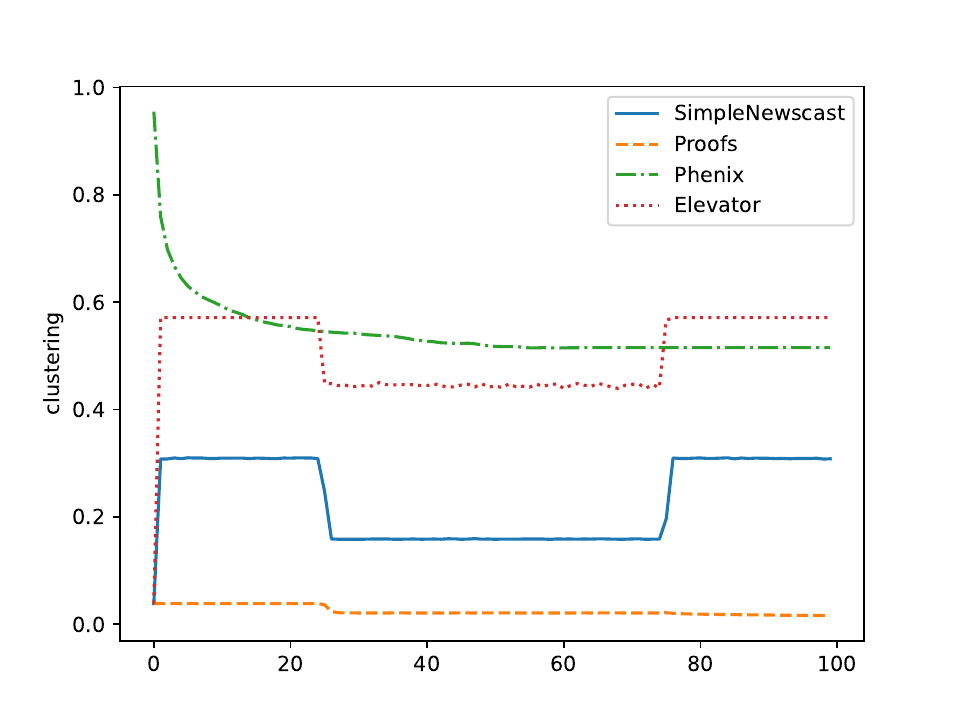}
    \caption{Clustering coefficient computed with churn, for each algorithm, every 10 cycles}
    \label{fig:ClustCoefChurn}
\end{figure}

\begin{figure}[H]
    \includegraphics[width=0.85\columnwidth]{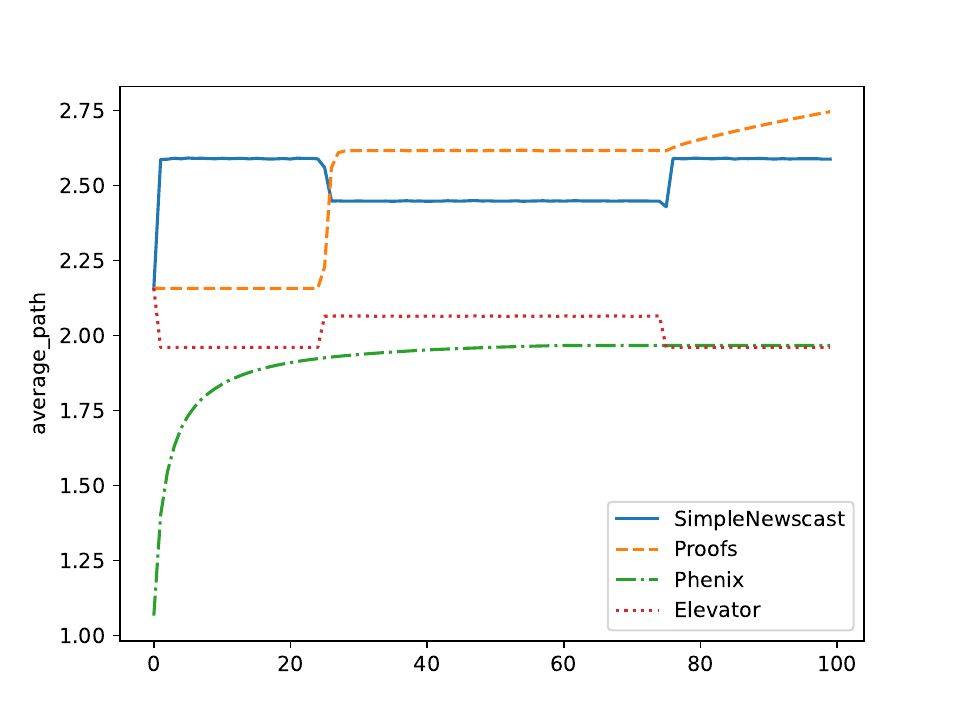}
    \caption{Average path length computed with churn, for each algorithm, every 10 cycles}
    \label{fig:AveragePathLengthChurn}
\end{figure}

\begin{figure}[H]
    \includegraphics[width=0.85\columnwidth]{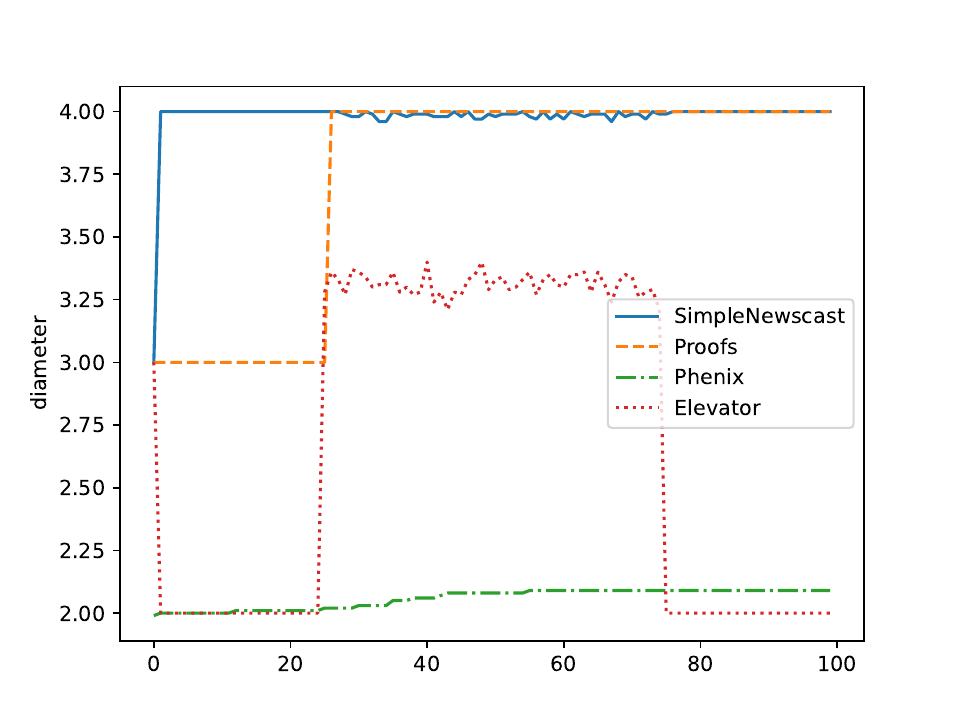}
    \caption{Diameter computed with churn, for each algorithm, every 10 cycles}
    \label{fig:DiameterChurn}
\end{figure}

The in-degree distribution of Elevator remains the same, with 10 hubs. PROOFS seems affected by churn, as the mean degree distribution goes to 10 instead of 20 without churn. In Figure~\ref{fig:ClustCoefChurn} we can observe that we have almost the same results as the results obtained without churn for the clustering coefficient. For the average path length, PROOFS is the most affected, with a value going from 2.25 without churn to a value of 2.5 with churn, and the value keep increasing after the end of the churn, going up to 2.75, as we can see in Figure~\ref{fig:AveragePathLengthChurn}. In Figure~\ref{fig:DiameterChurn}, we can see that the diameter varies with churn, with a mean going up to 3.25 instead of 2.0, but the values for Phenix and Elevator remain below the ones of Newscast and PROOFS.

\subsection{Resilience to hub-targeted attacks} 

We hereby analyze the performance of the four algorithms after a targeted attack on the hubs during the execution of the simulation. To simulate a hub-targeted attack, we disconnected 10 nodes that have the highest in-degree in the middle of the simulated scenario.

\begin{figure}[H]
    \includegraphics[width=0.85\columnwidth]{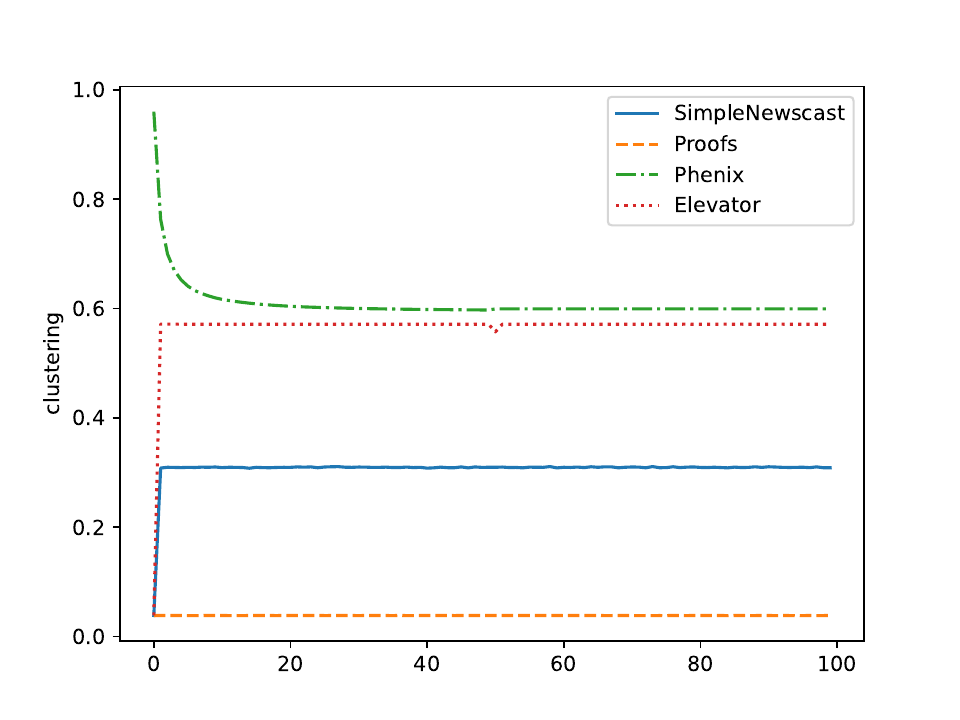}
    \caption{Clustering coefficient computed with a hub-targeted attack, for each algorithm, every 10 cycles}
    \label{fig:ClustCoefCrashHub}
\end{figure}

\begin{figure}[H]
    \includegraphics[width=0.85\columnwidth]{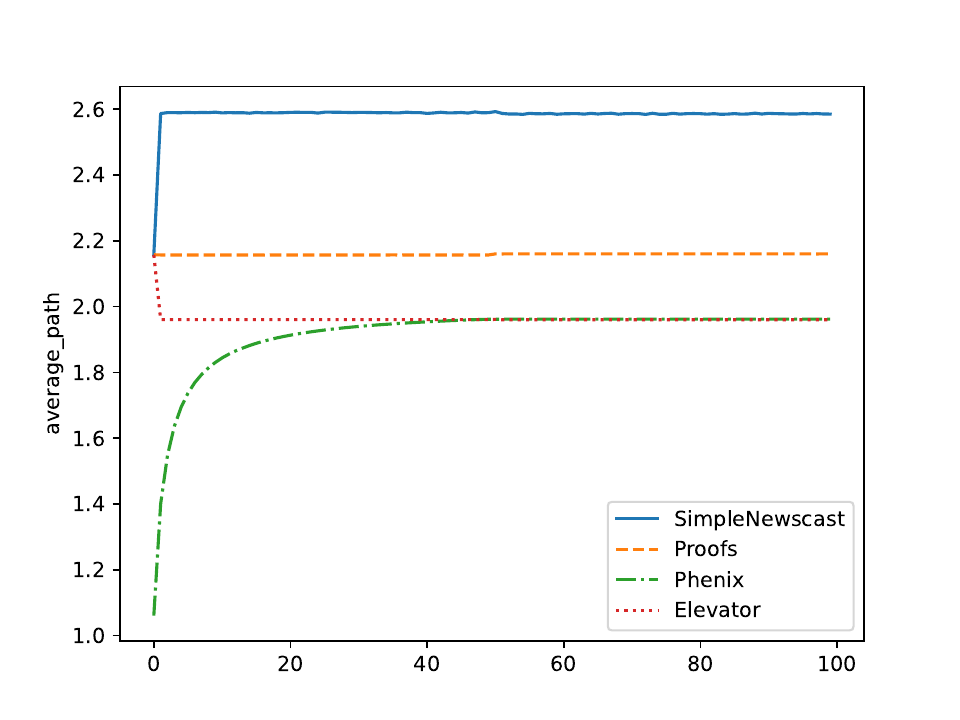}
    \caption{Average path length computed with a hub-targeted attack, for each algorithm, every 10 cycles}
    \label{fig:AveragePathLengthCrashHub}
\end{figure}

\begin{figure}[H]
    \includegraphics[width=0.85\columnwidth]{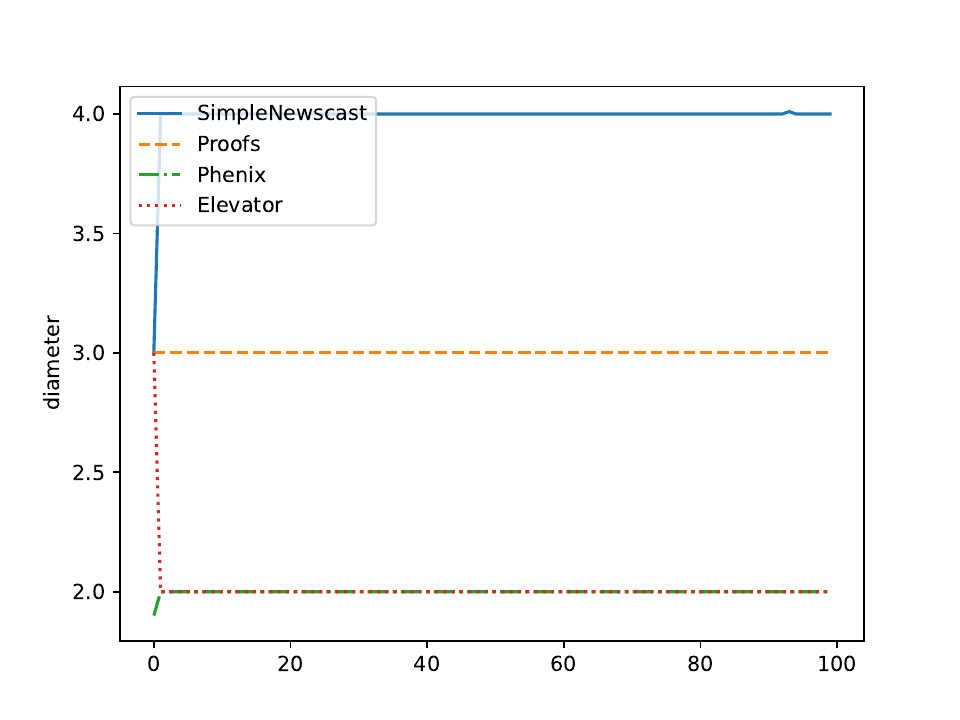}
    \caption{Diameter computed with a hub-targeted attack, for each algorithm, every 10 cycles}
    \label{fig:DiameterCrashHub}
\end{figure}

Logically, Newscast and PROOFS are not affected by the attack, as there are no hubs in the networks built by these algorithms. 
For Elevator, the in-degree distribution remains similar, with 10 high-in-degree peers that have each an in-degree of 989. 
We are thus confident in the capacity of our algorithm to promote new nodes to the position of hubs if the previous hubs were disconnected. In Figure~\ref{fig:ClustCoefCrashHub} we can see that we have almost the same results as the results obtained without crashes for the clustering coefficient. Its the same for the average path length and the diameter, there is no impact, as we can see in Figure \ref{fig:AveragePathLengthCrashHub} and \ref{fig:DiameterCrashHub}.

\subsection{Summary.} 
\begin{figure}[H]
    \includegraphics[width=0.85\columnwidth]{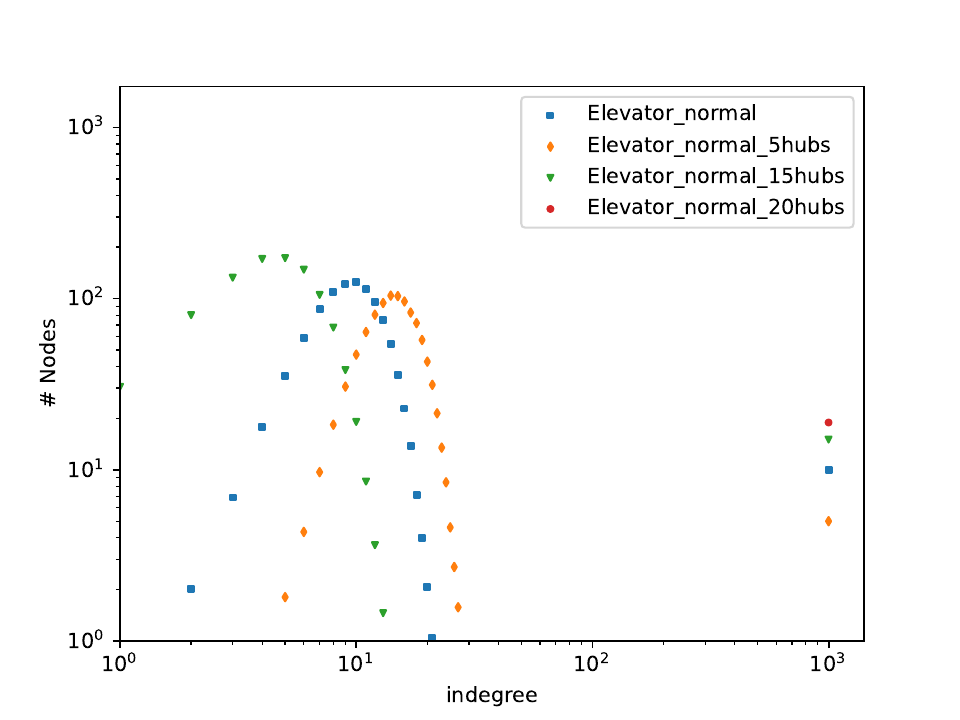}
    \caption{In-degree distribution of the network, after the run of the Elevator algorithm, with a variable number of hubs (5 hubs, 10 hubs, 15 hubs, 20 hubs), no failures.}
    \label{fig:degreeDistributionVariableNbHubs}
\end{figure}

\begin{figure}[H]
    \includegraphics[width=0.85\columnwidth]{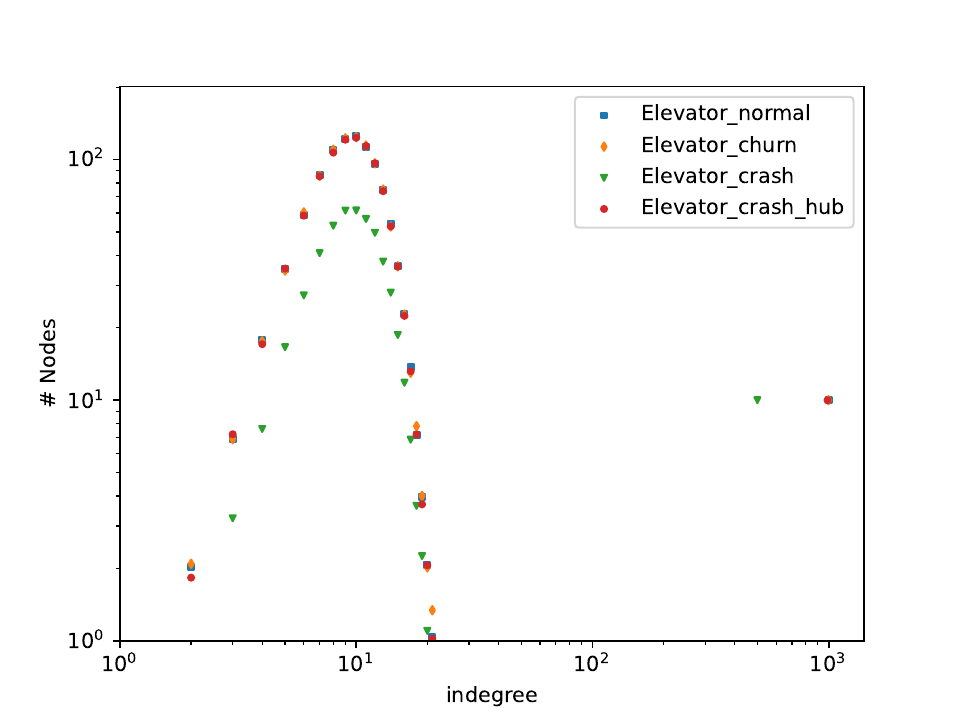}
    \caption{In-degree distribution of the network, after the run of the Elevator algorithm, during each context (no failures, 50\% crash, churn, and hub-targeted attack).}
    \label{fig:CompareContext}
\end{figure}

\begin{figure}[H]
    \includegraphics[width=0.85\columnwidth]{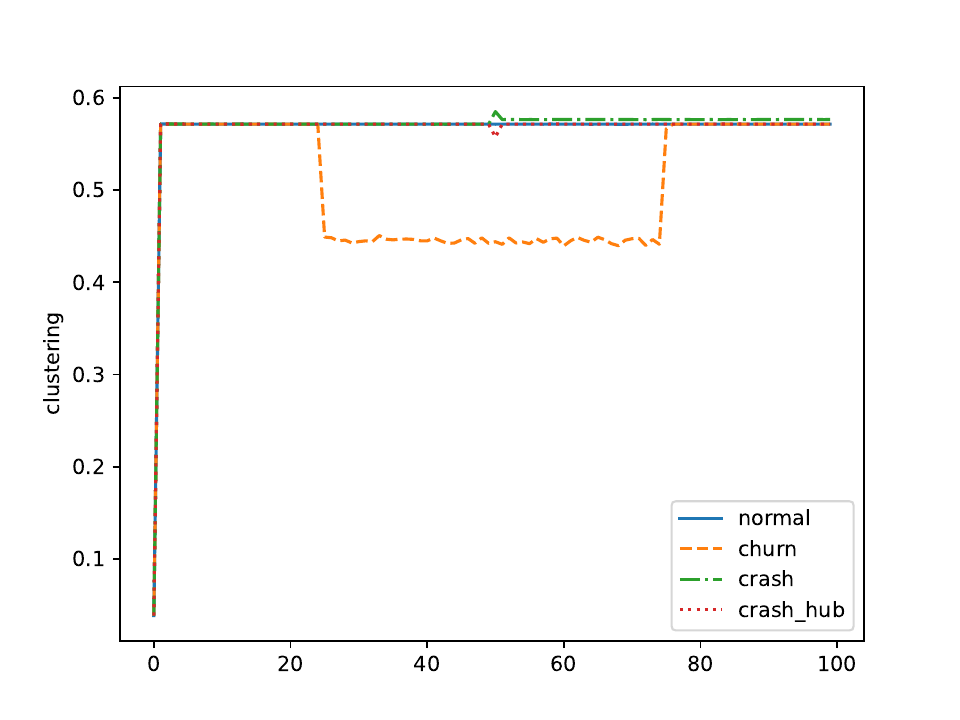}
    \caption{Clustering of the network, after the run of the Elevator algorithm, during each context (no failures, 50\% crash, churn, and hub-targeted attack).}
    \label{fig:ElevatorContextCoefClust}
\end{figure}

\begin{figure}[H]
    \includegraphics[width=0.85\columnwidth]{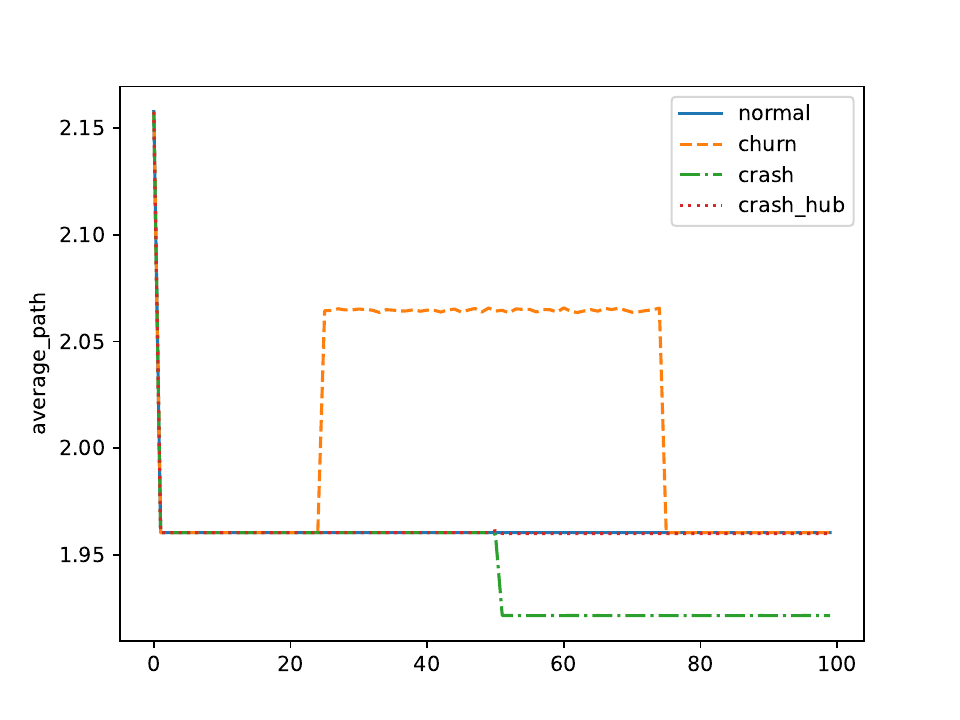}
    \caption{Average path length of the network, after the run of the Elevator algorithm, during each context (no failures, 50\% crash, churn, and hub-targeted attack).}
    \label{fig:ElevatorAveragePathLength}
\end{figure}

\begin{figure}[H]
    \includegraphics[width=0.85\columnwidth]{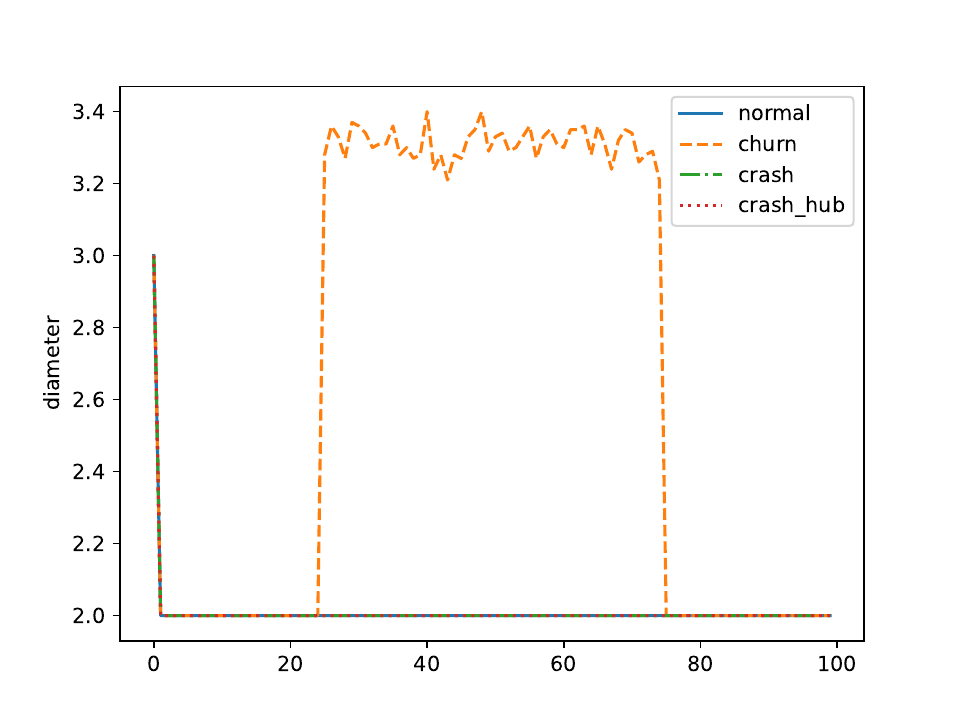}
    \caption{Diameter of the network, after the run of the Elevator algorithm, during each context (no failures, 50\% crash, churn, and hub-targeted attack).}
    \label{fig:ElevatorDiameter}
\end{figure}

We have compared the in-degree distribution of the network after the run of the Elevator algorithm for a various number of hubs in Figure~\ref{fig:degreeDistributionVariableNbHubs}, and also for each context of simulation in Figure~\ref{fig:CompareContext}. The shape of the degree distribution remains consistent across different hub counts, except for a scenario with 20 hubs where nodes exclusively connect to these hubs (resulting in a multi-star topology). This phenomenon aligns with the prescribed number of preferred connections (\emph{h} = \emph{c} = 20), where nodes exclusively link to elevated hub nodes, omitting random connections entirely. The shape of distribution also remains consistent across failure contexts. In Figure~\ref{fig:ElevatorContextCoefClust}, \ref{fig:ElevatorAveragePathLength} and \ref{fig:ElevatorDiameter}, we compare Elevator across all contexts for the different metrics, and we can see that there are not many variations in values, as expected from the definition of our protocol and as seen in previous comparative analyses presented above. Another notable feature is that Elevator seems more stable than Phenix. This is because once the hubs are in place they do not change (except in the event of failures), which provides stability in terms of network diameter or average path length.

\FloatBarrier

\section{Conclusion}
\label{sec:5}
We proposed a novel peer sampling algorithm, Elevator, designed for unstructured P2P networks, which facilitates the organic promotion of specific nodes to serve as hubs. Our simulations confirm that the Elevator algorithm successfully maintains network connectivity, constructs networks with low diameters, achieves stability with a defined number of hubs (denoted as \emph{h}), and demonstrates resilience against crashes, churn, and targeted attacks on hubs.
The distinctive aspect of our work lies in our pursuit of developing an unstructured network model with inherent hub nodes. 
We anticipate that this work will pave the way for a new category of algorithms known as "hub sampling algorithms", which could hold significant relevance for specific decentralized applications. For instance, such algorithms may accelerate the transmission of machine learning models in federated learning scenarios or automate the selection of validators in blockchain networks, thus potentially replacing the need for traditional proof-of-work protocols.
While our current study does not delve into these specific use cases, we envision exploring federated learning applications within this network paradigm in future investigations.

\bibliography{ref}{}
\bibliographystyle{plain}

\newpage
\appendices

\section{Pseudo-code of Elevator Algorithm}
\label{sec:algoElevator}
In this section we propose the detailed pseudo-code of Elevator algorithm presented in Algorithm~\ref{algo:algoLabel} and Algorithm~\ref{algo:algoLabelBackground}.  

\begin{figure}[H]
\removelatexerror
\begin{algorithm}[H]
  \caption{Elevator Algorithm (active thread)}
  \label{algo:algoLabel}
  \KwData{initial peer list: \emph{cache}}
  \KwData{cache size: \emph{c}}
  \KwData{number of hubs desired: \emph{h}}
  \KwData{initial backward list: \emph{$\mathit{backward\_peers}$} (empty)}
  \Loop{
    \For{$\mathit{peer}\in \mathit{backward\_peers}$}{	
  	  \If{peer not responding}
  	  {$\mathit{backward\_peers.remove(peer)}$}
  	}
    \For{$\mathit{peer}\in \mathit{cache}$}{
  	  \If{peer not responding}
  	  {$\mathit{cache.remove(peer)}$}
  	}
    $\mathit{frequency\_{map}} \gets \{\}$\\ 
    \For{$\mathit{peer}\in \mathit{cache}$}{
        $\mathit{peer\_cache} \gets \mathit{send(CACHE\_{REQUEST},peer)}$\\
        $\mathit{frequency\_{map}} \gets \mathit{frequency\_{map}} \cup \mathit{peer\_cache}$\\
  	}
    $\mathit{preferred} \gets \mathit{frequency\_{map}.sortByFrequency().select(number=c)}$\\    
    $\mathit{frequency\_{map}.remove(preferred)}$\\
    $\mathit{preferred\_backward} \gets \{\}$\\ 
    \For{$\mathit{peer}\in \mathit{preferred}$}{
        $\mathit{peer\_backward\_peers} \gets \mathit{send(BACKWARD\_{REQUEST},peer)}$\\
        $\mathit{preferred\_backward} \gets \mathit{preferred\_backward} \cup \mathit{peer\_backward\_peers}$\\
  	}
    $\mathit{preferred\_backward.shuffle()}$\\
    $\mathit{cache} \gets \{\}$\\
    $\mathit{cache} \gets \mathit{preferred}[1..h] + \mathit{preferred\_backward}[1..c-h]$\\
    \While{$\mathit{cache.size()} < c$}{
        $\mathit{peer} \gets \mathit{frequency\_{map}.selectRandom()}$ \\
        $\mathit{cache.append(peer)}$\\
    }
  }
      
\end{algorithm} 
\end{figure}

\begin{figure}[H]
\removelatexerror
\begin{algorithm}[H]
  \caption{Elevator Algorithm (background thread)}
  \label{algo:algoLabelBackground}
  \KwData{max number of backward connections to send: \emph{$\mathit{maxsize\_buffer\_backward}$}}
  \Loop{
    $\mathit{request, peer} \gets \mathit{receive()}$\\
    \If{$\mathit{request} = \mathit{CACHE\_{REQUEST}}$}{
        $ \mathit{send(cache, peer)}$\\
        $ \mathit{backward\_peers.add(peer)}$\\
    }
    \If{$\mathit{request} = \mathit{BACKWARD\_{REQUEST}}$}{
        $ \mathit{backward\_peers.shuffle()}$\\
        $ \mathit{send(backward\_peers}[:\mathit{maxsize\_buffer\_backward}], \mathit{peer})$\\
    }
  }
      
\end{algorithm}
\end{figure}

\section{Theoretical analysis}
\label{sec:3}

Assume the network initially adopts a random regular graph topology with a total of \emph{n} nodes, where each node has outgoing connections to \emph{c} other nodes. Following each iteration of the Elevator algorithm, the network, with high probability, establishes preferential attachment links.

For a node to establish preferential attachment links, it must have at least two neighbors that share at least one common neighbor. If all neighbors of a node have distinct sets of neighbors, the node cannot select the most frequent 2-distance neighbors as preferred nodes. The probability that a node does not share any neighbors with another node is $1 - \frac{c}{n}$. Given that each node has \emph{c} neighbors, the probability that none of its neighbors share neighbors is approximately equal to $(1 - \frac{c}{n})^{\binom{c}{2}}$. Even if preferential links are not created during the first iteration of the algorithm for a given node, the algorithm still establishes random links. When the algorithm is run a second time on the node, the probability of not creating preferential links remains the same. Therefore, for a single node, the probability of not creating preferential links after \emph{t} iterations of the algorithm is approximately equal to $((1 - \frac{c}{n})^{\binom{c}{2}})^t$, assuming stochastic independence between iterations.

This gives us a reasonable approximation of the creation of preferential links. For $n = 1000$ and $c = 20$, this probability is $(1 - \frac{20}{1000})^{\binom{20}{2}} \approx 2.5\%$. After 20 iterations of the algorithm, the probability becomes $(2.5\%)^{20} \approx 10^{-34}$.

Let \emph{S} be a configuration of the system with \emph{h} defined hubs (each with an in-degree equal to $n-1$). The analysis of the convergence towards such configurations is outside the scope of this article, but our simulations practically show that they are indeed reached. Furthermore, it has been shown in the literature that power-law networks (with hubs) can be formed using the principle of preferential attachment~\cite{lynn2024emergent}.

Now, starting from a configuration satisfying \emph{S}, the network maintains this property after each iteration of the Elevator algorithm (on one node) with high probability. Indeed, when a node \emph{i} executes the Elevator algorithm, it selects \emph{h} nodes that are 2-distance neighbors and have the highest in-degree. Since we start with \emph{h} defined hubs, with all other nodes connected to them, node \emph{i} always chooses these hubs as preferred nodes, and maintains connections to them. The only way for another node to be selected as a preferred node is for it to have all neighbors of node \emph{i} connected to it.

Assuming that each node maintains \emph{h} connections to the hubs and $c-h$ random connections, the probability of a node being a (random-connected) neighbor of another node is $\frac{c-h}{n}$, so the probability for each neighbor of node \emph{i} to share the same (random-connected) neighbor is $(\frac{c-h}{n})^c$. If we suppose that node \emph{i} selects \emph{h} nodes to connect to using preferential attachment, the probability of the node connecting to this newly selected node is $\frac{h}{h+1}$. Therefore, the final probability that the network remains in a configuration satisfying \emph{S} is $1 - \frac{h \cdot (\frac{c-h}{n})^c}{h+1}$.
For $n = 1000$, $c = 20$, and $h = 10$, this probability is $1 - \frac{10 \cdot (\frac{1}{100})^{20}}{11} \approx 1 - 10^{-39}$. 
To enforce a probability of 1 for maintaining the same hubs, we can adapt our algorithm to avoid selecting any new hubs if the previous hubs are still connected.

When the network is in a configuration satisfying \emph{S}, and assuming there are no failures, the network remains connected (in an undirected manner) with high probability after each run of the algorithm. This is because each node running the Elevator algorithm, with high probability, remains in a configuration satisfying \emph{S} (with the same hubs). Since the hubs are connected to all the nodes, the network remains connected. Furthermore, even if the risk of disconnection is exponentially low, we can incorporate a disconnection detection mechanism and allow a node to re-initiate its connection to the network using the \emph{init} method if a disconnection is detected.

\section{Description of PROOFS, Newscast and Phenix algorithms}
\label{sec:algorithms}

As our goal is to present our new hub sampling algorithm and compare it to previous peer sampling algorithms, we will (briefly) present three peer sampling algorithms (PROOFS, Newscast, and Phenix). 

\subsection{The PROOFS algorithm:} The PROOFS algorithm, as presented in~\cite{stavrou2004lightweight} is a very simple algorithm used to create a peer sampling service. At each cycle, each node initiates a neighbor exchange (or shuffling) with another peer \emph{q} chosen at random. The peer selects a random subset of size \emph{l} (the shuffle length, a global parameter) and sends this subset to \emph{q}. Upon reception of the subset, the node \emph{q} also selects a random subset and sends it to \emph{p}. When the node receives the subset of \emph{q}, it replaces the previous entry in its cache, starting with the empty cache slots (if any) and then replacing entries previously sent to \emph{q}. The parameters of the algorithm are \emph{c}, the size of the list of outgoing connections, and \emph{l}, the shuffle length, i.e. the number of outgoing connections exchanged with a peer during a neighbor exchange. The cache list is implemented as an array of size \emph{c}. The list is initialized with random values (random connections to other nodes of the network).

\begin{algorithm}
  \caption{PROOFS algorithm (active thread)}
  \KwData{initial peer list: \emph{cache}}
  \KwData{cache size: \emph{c}}
  \KwData{shuffle length: \emph{l}}
  \KwData{node address: \emph{p}}
  \Loop{
    $\mathit{subset} \gets \mathit{selectRandomSubset(cache, l)}$\\ 
    $q \gets \mathit{selectRandom(subset)}$\\ 
    $\mathit{subset.remove(q)}$\\
    $\mathit{subset.add(p)}$\\
    $\mathit{send(q, subset)}$\\
    $\mathit{subset}_q \gets \mathit{receive(q)}$\\ 
    $\mathit{subset}_q.\mathit{remove(p)}$\\
    $\mathit{subset}_q.\mathit{removeAll(cache)}$\\
    $\mathit{cache} \gets \mathit{subset}_q$\\ 
  }
\label{algoLabelProofs}
\end{algorithm}

\begin{algorithm}
  \caption{PROOFS algorithm (background thread)}
  \KwData{peer list: cache}
  \KwData{shuffle length: l}
  \KwData{node address: p}
  \Loop{
    $q, \mathit{subset}_q \gets \mathit{receive()}$\\ 
    $\mathit{subset} \gets \mathit{selectRandomSubset(cache, l)}$\\ 
    $\mathit{send(q, subset)}$\\
    $\mathit{subset}_q.\mathit{remove(p)}$\\
    $\mathit{subset}_q.\mathit{removeAll(cache)}$\\
    $\mathit{cache} \gets \mathit{subset}_q$\\ 
  }
\label{algoLabelProofsBack}
\end{algorithm}

\begin{figure}
\centering 
\begin{tikzpicture}[node distance={15mm}, thick, main/.style = {draw, circle}] 
\node[main, fill=blue!40] (1) {$1$};
\node[main] (2) [above right of=1] {$2$};
\node[main] (3) [below of=1] {$3$}; 
\node[main, fill=green!20] (4) [right of=1] {$4$};
\node[main] (5) [above right of=4] {$5$}; 
\node[main] (7) [left of=1] {$7$};
\node[main] (8) [below of=4] {$8$};
\node[main] (6) [right of=8] {$6$};
\draw[->] (1) -- (2);
\draw[->] (1) -- (3);
\draw[->] (1) -- (4);
\draw[->] (1) -- (7);
\draw[->] (4) -- (5);
\draw[->] (4) -- (6);
\draw[->] (4) -- (8);
\draw[->] (4) -- (3);
\end{tikzpicture}
\begin{tikzpicture}[node distance={15mm}, thick, main/.style = {draw, circle}] 
\node[main, fill=blue!40] (1) {$1$};
\node[main] (2) [above right of=1] {$2$};
\node[main] (3) [below of=1] {$3$}; 
\node[main, fill=green!20] (4) [right of=1] {$4$};
\node[main] (5) [above right of=4] {$5$}; 
\node[main] (7) [left of=1] {$7$};
\node[main] (8) [below of=4] {$8$};
\node[main] (6) [right of=8] {$6$};
\draw[->] (4) -- (2);
\draw[->] (4) -- (1);
\draw[->] (1) -- (7);
\draw[->] (1) -- (5);
\draw[->] (1) -- (6);
\draw[->] (1) -- (8);
\draw[->] (4) -- (3);
\end{tikzpicture}
\caption{Before and after the execution of the PROOFS algorithm. The node 1 sends its address alongside the address of 2 and 3 to the node 4. Node 4 sends back the addresses of nodes 5,6 and 8. Node 1 replaces the connection to 2, 3, and 4 with connections to 5,6 and 8. Node 4 drops connection to 5,6, and 8 and connect to nodes 2 and 1 (it is already connected to 3).} \label{fig:proofsExample}  
\end{figure}  

The goal of the algorithm is to produce a network that is “well-mixed”, in the sense that after enough shuffling operations, the node’s neighbors are essentially drawn at random from the set of all peers.

\subsection{The Newscast algorithm:} The Newscast algorithm\cite{jelasity2007gossip} is similar to PROOFS but is more generic and adds the idea of "age" for the node descriptors. The age of the node descriptors is incremented at each cycle. The goal is to create a peer sampling service that allows each node of the network to connect to a random subset of the nodes in the network. The parameters of the algorithm are \emph{c}, the size of the list of outgoing connections, the mode of the peer selection (random or tail, but for simulations we only used random), and the mode of view propagation. 
The mode of view propagation can be push, pull, or push-pull, as described below: 
    \begin{itemize}
        \item Push strategy: At each cycle, a node will send its knowledge to the selected node
        \item Pull strategy: At each cycle, a node will ask for knowledge from the selected node and wait for the answer.
        \item Push-Pull strategy: At each cycle, a node will both use a Push and a Pull strategy.    
    \end{itemize}
As the push-pull mode is the most efficient\cite{jelasity2007gossip}, this is the mode we will present and the one that we used in our simulations. The cache list is implemented as an array of 2-tuple of size \emph{c}. The two elements of the tuple are the node descriptor and the associated age of the descriptor. The list is initialized with random values for the node descriptors (random connections to other nodes of the network) and with 0s for the age of the node descriptors.
At each cycle, each node initiates an exchange of membership information with a neighbor chosen at random. The node sends a buffer that contains \emph{$\frac{c}{2}-1$} random node descriptors from its cache to the other node, with \emph{c} the parameter representing the size of the cache. The other node replies to the message with a similar message also containing a buffer with \emph{$\frac{c}{2}-1$} nodes descriptors. The node then merges the received buffer with its cache and filters the elements (removing the duplicates, the older elements, and finally removing the sent elements) to achieve a cache of the same size as before. If there are still too many elements, the algorithm removes elements of the cache at random until the length of the \emph{cache} is \emph{c}.

\begin{algorithm}
  \caption{Newscast algorithm (active thread)}
  \KwData{initial peer list: cache}
  \KwData{cache size: \emph{c}}
  \KwData{node address: p}
  \Loop{
    $q \gets \mathit{selectRandom(cache)}$\\ 
    $\mathit{buffer} = \gets \{\}$\\
    $\mathit{buffer.append}({\mathit{address}=p,\mathit{age}=0})$\\
    $\mathit{cache.permute()}$\\
    $\mathit{buffer.append(view.head}(c/2-1))$\\
    $\mathit{send(q, buffer)}$\\
    $\mathit{buffer}_q \gets \mathit{receive(q)}$\\ 
    $\mathit{cache.append}(\mathit{buffer}_q)$\\
    $\mathit{cache.removeDuplicates()}$\\
    $\mathit{cache.removeOldItems()}$\\
    $\mathit{cache.removeHead()}$\\
    $\mathit{cache.removeRandom()}$\\
    $\mathit{cache.IncrementAllItemsAge()}$\\
  }
\label{algoLabelNewscast}
\end{algorithm}

\begin{algorithm}
  \caption{Newscast algorithm (background thread)}
  \KwData{peer list: cache}
  \KwData{node address: p}
  \Loop{
    $q,\mathit{buffer}_q \gets \mathit{receive()}$\\ 
    $\mathit{buffer} = \mathit{new buffer()}$\\
    $\mathit{buffer.append}({\mathit{address}=p,\mathit{age}=0})$\\
    $\mathit{cache.permute()}$\\
    $\mathit{buffer.append(view.head}(c/2-1))$\\
    $\mathit{send(q, buffer)}$\\
    $\mathit{cache.append(buffer_q)}$\\
    $\mathit{cache.removeDuplicates()}$\\
    $\mathit{cache.removeOldItems()}$\\
    $\mathit{cache.removeHead()}$\\
    $\mathit{cache.removeRandom()}$\\
    $\mathit{cache.IncrementAllItemsAge()}$\\
  }
\label{algoLabelNewscastBack}
\end{algorithm}

As with PROOFS, the Newscast algorithm allows the creation of a network that has the same behavior as a random graph. In particular, this allows the network to be very resilient to failures and churn, as explained in~\cite{jelasity2007gossip}.

\subsection{The Phenix algorithm:} The Phenix algorithm\cite{wouhaybi2004phenix} is a peer sampling algorithm that differs from PROOFS and Newscast in the sense that the goal of the authors is to create an algorithm that is both resilient to failures and with a low-diameter. The algorithm is inspired by the concept of preferential attachment\cite{barabasi2002evolution} and constructs a network with a topology that is close to a power-law. Contrary to the two previous algorithms, the Phenix algorithm only executes once for each node, when the node enters the network. The node splits its cache in 2 parts $G_{random}$ and $G_{friends}$. Then it connects directly to the nodes in $G_{random}$ and asks for the list of neighbors for each node in $G_{friend}$ and adds them to the list $G_{candidates}$. Each node in $G_{friend}$ also sends a ping message to all its neighbors and all neighbors add the new node to their $\Gamma$ list, to prevent crawling from malicious nodes. The new node then sorts this list of distance two neighbors ($G_{candidates}$) and selects the \emph{s} more frequent nodes and connects to them ($G_{preferred}$). When a node receives a connection request from a new node, it will increment its internal counter and create a backward connection with this node if its counter's value is greater than the \emph{$\gamma$} constant. The parameters of the algorithm are \emph{c}, the number of outgoing connections, \emph{$\tau$}, the number of cycles a node is kept in the gamma list (fixed at 10), \emph{$\gamma$}, the constant limiting the number of backward connections (fixed to 20, thus a backward connection is created for 20 in-going connections) and \emph{s}, the number of preferential connections, chosen to the value of \emph{c/2} for our simulations. The cache list is implemented as an array of size \emph{c}. The list is initialized with random values (random connections to other nodes of the network). The  $\Gamma$ list is implemented as a linked list and initialized as an empty list.

\begin{algorithm}
  \caption{Phenix algorithm (active thread)}
  \KwData{cache size: \emph{c}}
  \KwData{initial peer list: cache}
  \KwData{initial backward list: $\mathit{backward\_peers}$ (empty)}
  \KwData{number of preferential connections: \emph{s}}
  $G_{\mathit{random}}, G_{\mathit{friend}} \gets \mathit{split(cache)}$\\ 
  $\mathit{cache} \gets \{\}$\\ 
  $\mathit{cache.append}(G_{\mathit{random}})$\\
  $G_{\mathit{candidates}} \gets \{\}$\\ 
  \For{$\mathit{peer}\in G_{\mathit{friend}}$}{
      $\mathit{neighbor\_list} \gets \mathit{send(peer, CACHE\_REQUEST)}$\\
      $G_{\mathit{candidates}} \gets G_{\mathit{candidates}} \cup \mathit{neighbor\_list}$\\
  }
  $\mathit{sort}(G_{\mathit{candidates}})$\\
  $G_{\mathit{preferred}} \gets G_{\mathit{candidates}}[0..(s-1)]$\\
  \For{$\mathit{peer}\in G_{\mathit{prefered}}$}{
      $\mathit{send(peer, CONNEXION\_REQUEST)}$\\
  }
  $\mathit{cache.append(G_{preferred})}$\\
\label{algoLabelPhenix}
\end{algorithm}

\begin{algorithm}
  \caption{Phenix algorithm (background thread)}
  \KwData{peer list: cache}
  \KwData{gamma list: $\Gamma$}
  \KwData{initial backward list: $\mathit{backward\_peers}$ (empty)}
  \KwData{number of preferential connections: s (fixed at c/2)}
  \KwData{internal counter for backward connections: $c_m$ (start at 0)}
  \KwData{backward connections constant: $\gamma$ (fixed at 20)}
\Loop{
    $\Gamma.\mathit{removeOldItems()}$\\
    $\mathit{request, peer} \gets \mathit{receive()}$\\
    \If{$\mathit{request} = \mathit{CACHE\_{REQUEST}}$}{
        $\mathit{send(cache, peer})$\\
        \For{$\mathit{node} \in \mathit{cache}$}{
            $\mathit{send(node, PING\_{REQUEST})}$\\
        }
    }
    \If{$\mathit{request} = \mathit{PING\_REQUEST}$}{
       $\Gamma.\mathit{add(peer)}$\\        
    }
    \If{$\mathit{request} = \mathit{CONNEXION\_REQUEST}$}{
        $c_m++$\\
        \If{$c_m >= \gamma$}{
            $ \mathit{backward\_peers.add(peer)}$\\
            $c_m \gets c_m - \gamma$\\
        }
    }
}
\label{algoLabelPhenixBack}
\end{algorithm}

To allow for the idea of preferential attachment to work, the network needs to be initialized with a small number of nodes (the number is set to 20 in~\cite{wouhaybi2004phenix} and we have chosen the same value for our simulations) and nodes are added progressively, with the number of nodes added at each cycle is drawn from a normal distribution $\mathcal{N}(2,1)$. The constructed network is a scale-free network, with a topology following a power law.

\section{Metrics used: Degree Distribution, Clustering, Average Path Length and Diameter}
\label{sec:metrics}

\subsection{Degree distribution}

The \emph{indegree (resp outdegree) distribution} of a network represents the probability distribution of these indegrees (resp outdegrees) over the whole network.   
\begin{itemize}
    \item A network that follows a random graph distribution (Erdős–Rényi model) should have a degree distribution that follows the probability $P(k) = \binom{n-1}{k} p^k (1-p)^{n-1-k}$
    \item A network that follows a power law (Barabási-Albert model) should have a degree distribution that follows the following probability $P(k) = Ck^{-\gamma}$
\end{itemize}
Indeed, observing the degree distribution should tell us if our algorithm creates a network with a topology closer to a random graph or one closer to a power-law.

\subsection{Clustering coefficient}
A random graph tends to have a low clustering coefficient, and a network with a lot of hubs will have a higher clustering coefficient.
The \emph{clustering coefficient} of a node is the number of edges between the neighbors of the node divided by the number of all possible edges between those neighbors. Intuitively, we can think of this coefficient as the measure of the degree to which nodes in a graph tend to cluster together (neighbors of the node are also neighbors of each other).
\[ C_i = \frac{2e_i}{k_i(k_i - 1)} \]

Where:
\begin{itemize}
  \item \( e_i \) is the number of closed triangles containing node \( i \).
  \item \( k_i \) is the degree of node \( i \), which is the number of links (edges) connected to that node.
\end{itemize}

\[ C = \frac{1}{n} \sum_{i=1}^{n} C_i \]

\subsection{Average Path Length}
As our goal is to have an algorithm that constructs a network that disseminates information in the network, our algorithm must produce a network topology with a low average path length. The average path length of each node was computed using the Floyd–Warshall algorithm.

The average path length is the average of the shortest path lengths over
all pairs of nodes in the graph.

\[a =\sum_{\substack{s,t \in V \\ s\neq t}} \frac{d(s, t)}{n(n-1)}\]

\subsection{Diameter}
The diameter of a graph is a measure of the longest distance between any two vertices (nodes) in the graph, measured in terms of the number of edges. In other words, the diameter of a graph is the maximum shortest path between any pair of nodes in the network.

\[ \text{diam}(G) = \max_{u,v \in V} \text{d}(u, v) \]

While the average path length provides a basic measure of information dissemination efficiency in algorithms, it may overlook disparities in dissemination speed across different nodes within the network. An algorithm could potentially have a favorable average path length but still exhibit uneven dissemination speeds among nodes due to varying distances. Calculating the network's diameter, however, offers a more comprehensive assessment.

\section{Additional results from simulations}
\label{sec:figures}

\begin{figure}[H]
    \includegraphics[width=0.85\columnwidth]{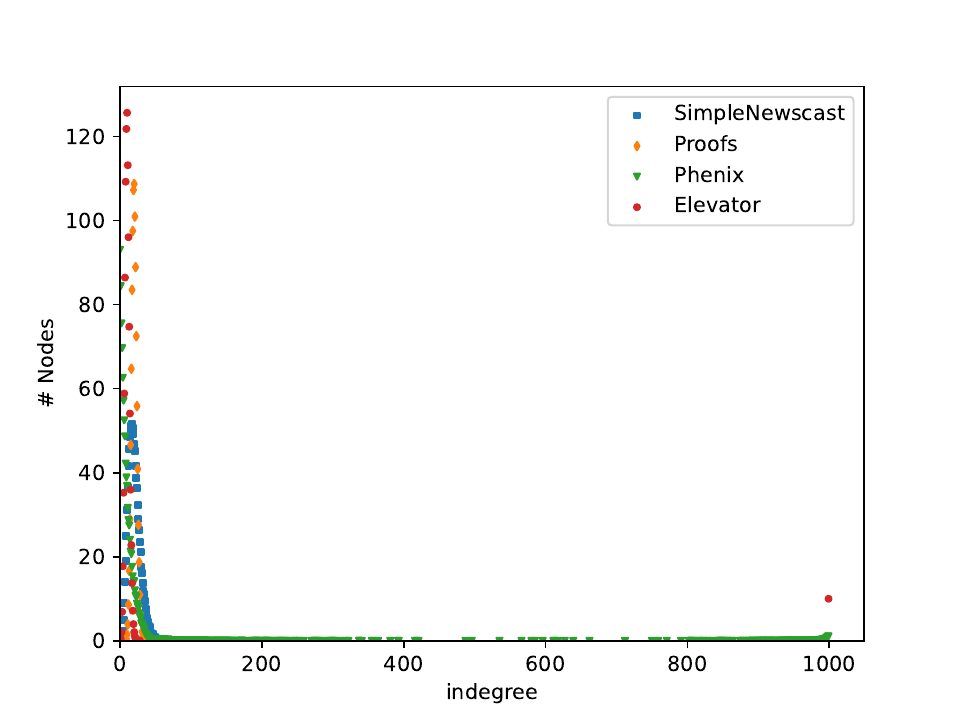}
    \caption{In-degree distribution of the network, after the run of the algorithm for 1000 cycles, for each algorithm.}
    \label{fig:degreeDistNormal}
\end{figure}

\begin{figure}[H]
    \includegraphics[width=0.85\columnwidth]{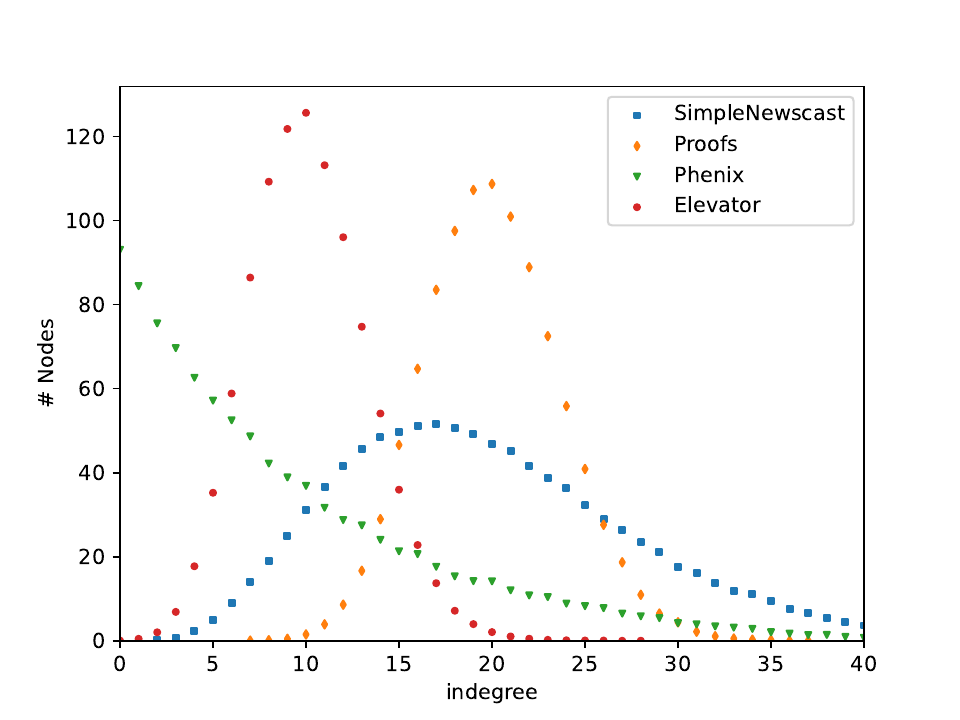}
    \caption{In-degree distribution of the network, after the run of the algorithm for 1000 cycles, for each algorithm, zoom on the beginning of distribution.}
    \label{fig:degreeDistNormalZoom}
\end{figure}
    
\begin{figure}[H]
    \includegraphics[width=0.85\columnwidth]{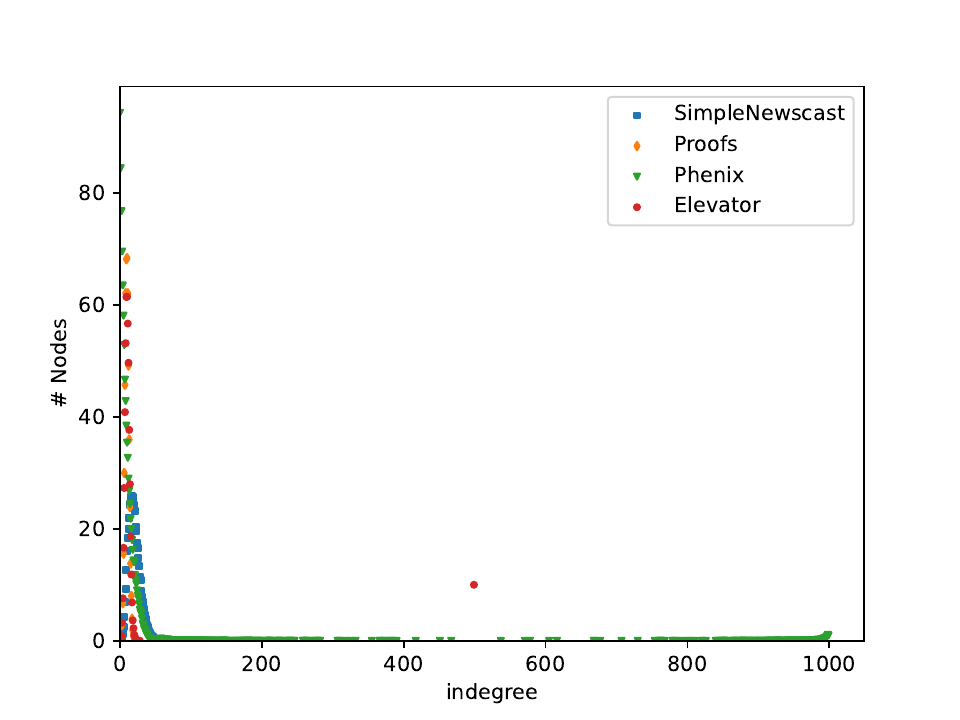}
    \caption{In-degree distribution of the network, with a 50\% crash, for each algorithm.}
    \label{fig:degreeDistCrash}
\end{figure}

\begin{figure}[H]
    \includegraphics[width=0.85\columnwidth]{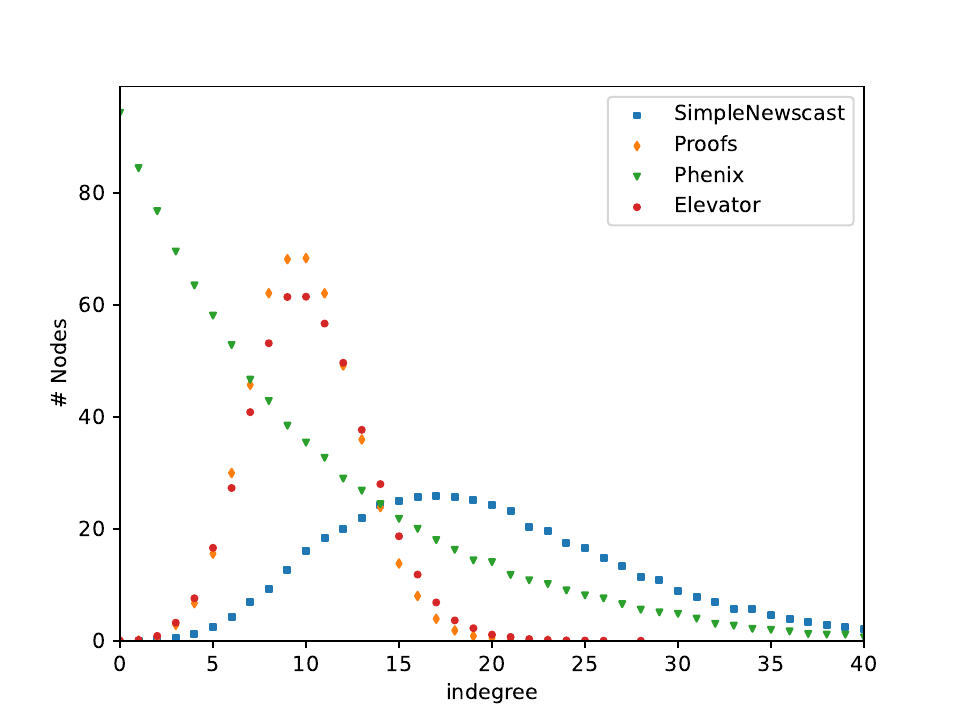}
    \caption{In-degree distribution of the network, with a 50\% crash, for each algorithm, zoom on the beginning of distribution.}
    \label{fig:degreeDistCrashZoom}
\end{figure}

\begin{figure}[H]
    \includegraphics[width=0.85\columnwidth]{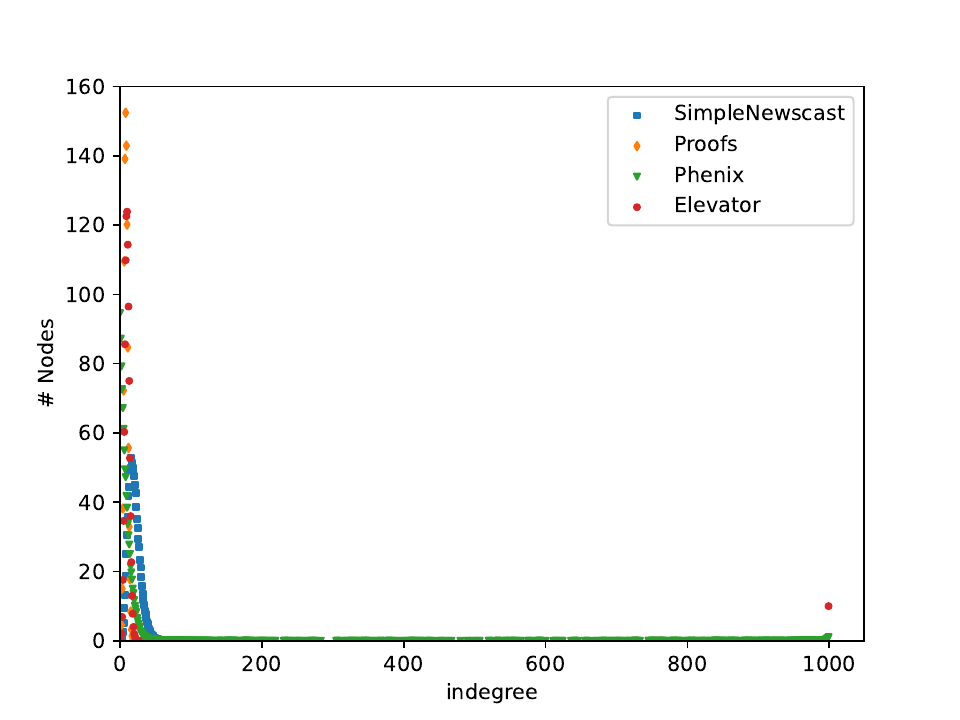}
    \caption{In-degree distribution of the network, with churn, for each algorithm.}
    \label{fig:degreeDistChurn}
\end{figure}

\begin{figure}[H]
    \includegraphics[width=0.85\columnwidth]{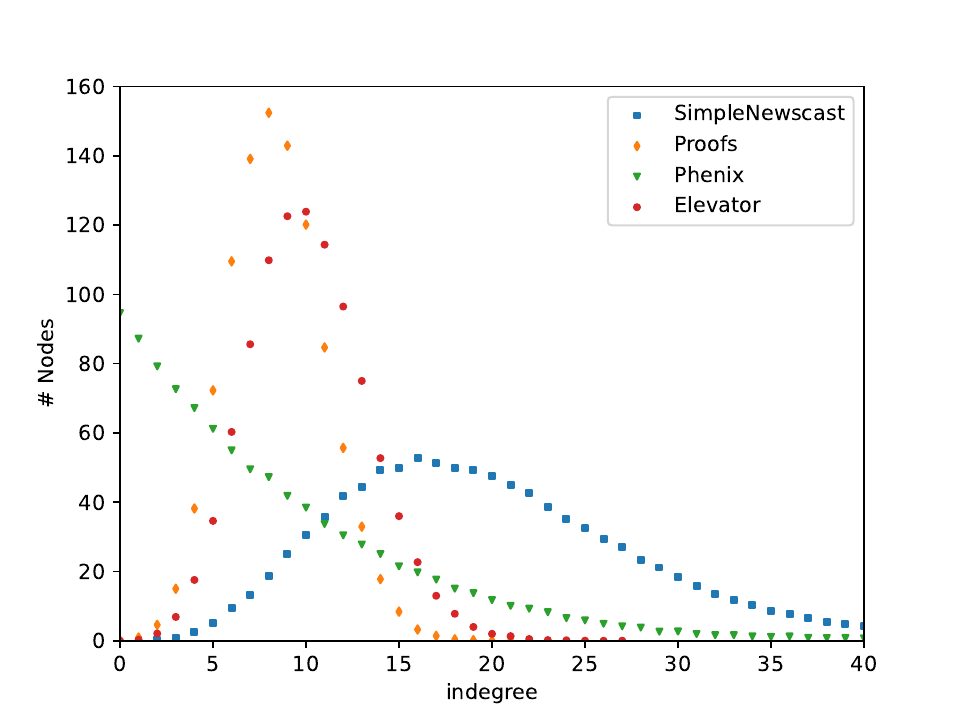}
    \caption{In-degree distribution of the network, with churn, for each algorithm, zoom on the beginning of distribution.}
    \label{fig:degreeDistZoomChurn}
\end{figure}

\begin{figure}[H]
    \includegraphics[width=0.85\columnwidth]{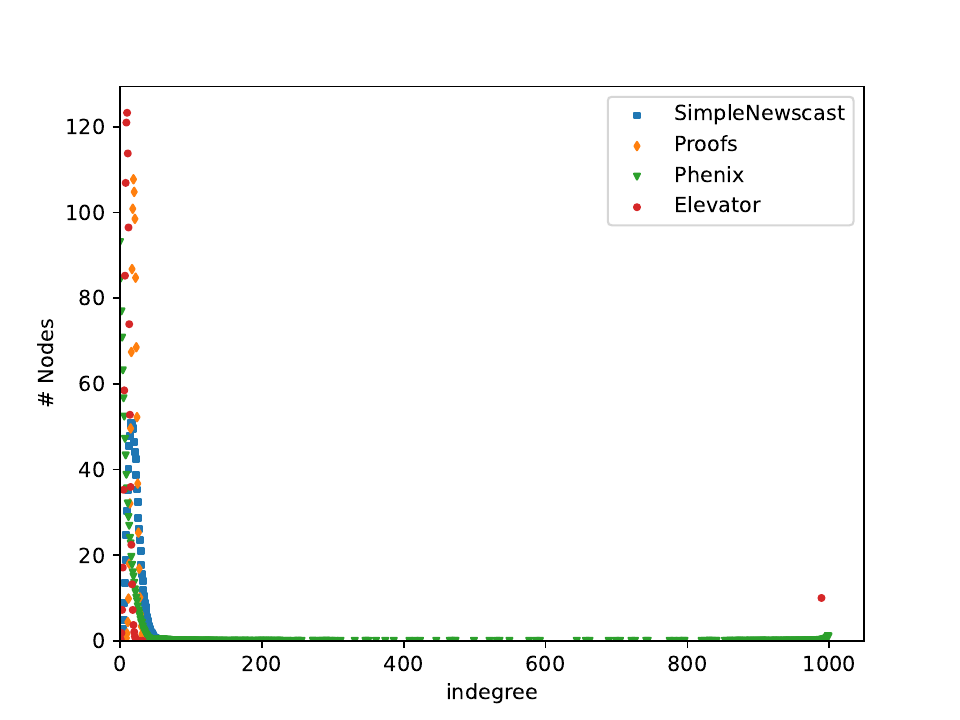}
    \caption{In-degree distribution of the network, with a hub-targeted attack, for each algorithm.}
    \label{fig:degreeDistCrashHub}
\end{figure}

\begin{figure}[H]
    \includegraphics[width=0.85\columnwidth]{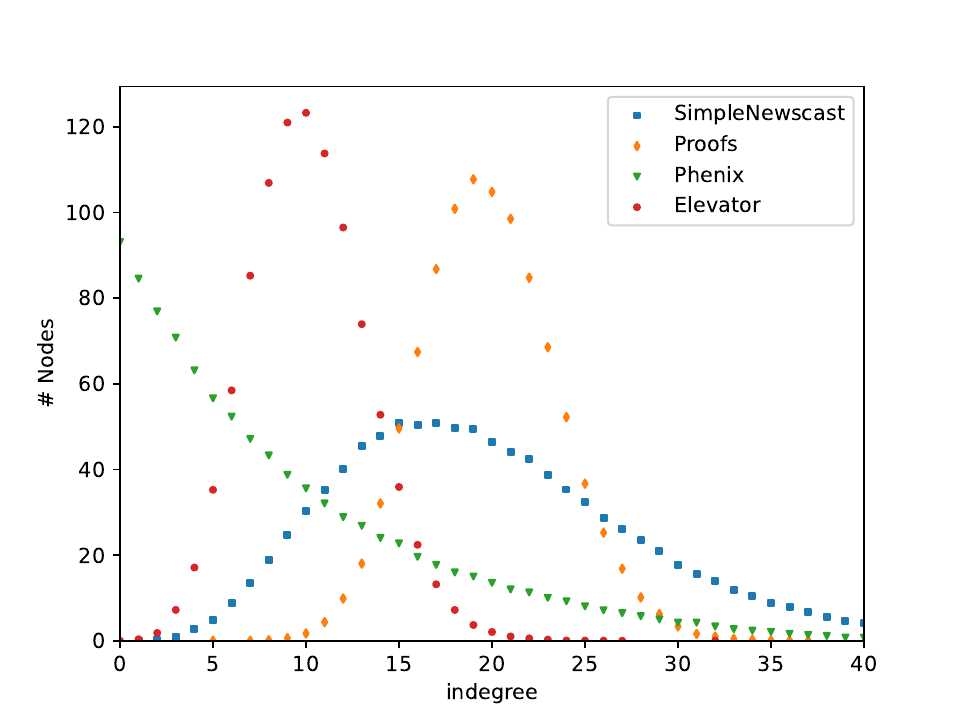}
    \caption{In-degree distribution of the network, with a hub-targeted attack, for each algorithm, zoom on the beginning of distribution.}
    \label{fig:degreeDistZoomCrashHub}
\end{figure}

\begin{figure}[H]
    \includegraphics[width=0.85\columnwidth]{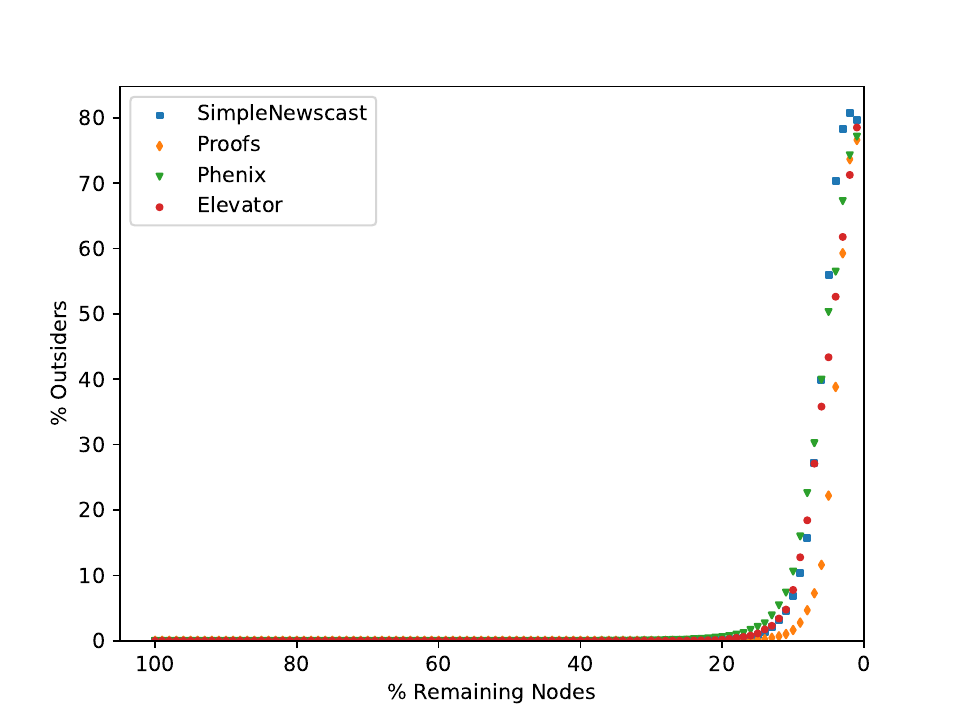}
    \caption{Analysis of the removal of all the nodes of the network one by one, and observing the number of nodes outside the biggest weakly connected cluster.}
    \label{fig:Component}
\end{figure}

\begin{figure}[H]
    \includegraphics[width=0.85\columnwidth]{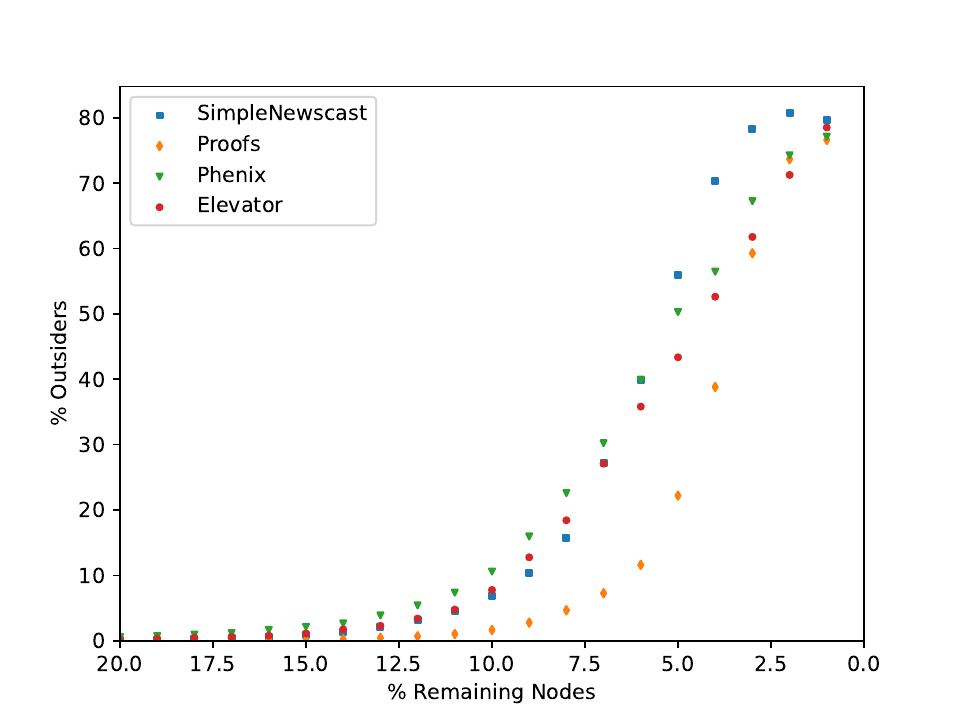}
    \caption{Analysis of the removal of all the nodes of the network one by one, zoom on the beginning of distribution.}
    \label{fig:ComponentZoom}
\end{figure}

\begin{figure}[H]
    \includegraphics[width=0.85\columnwidth]{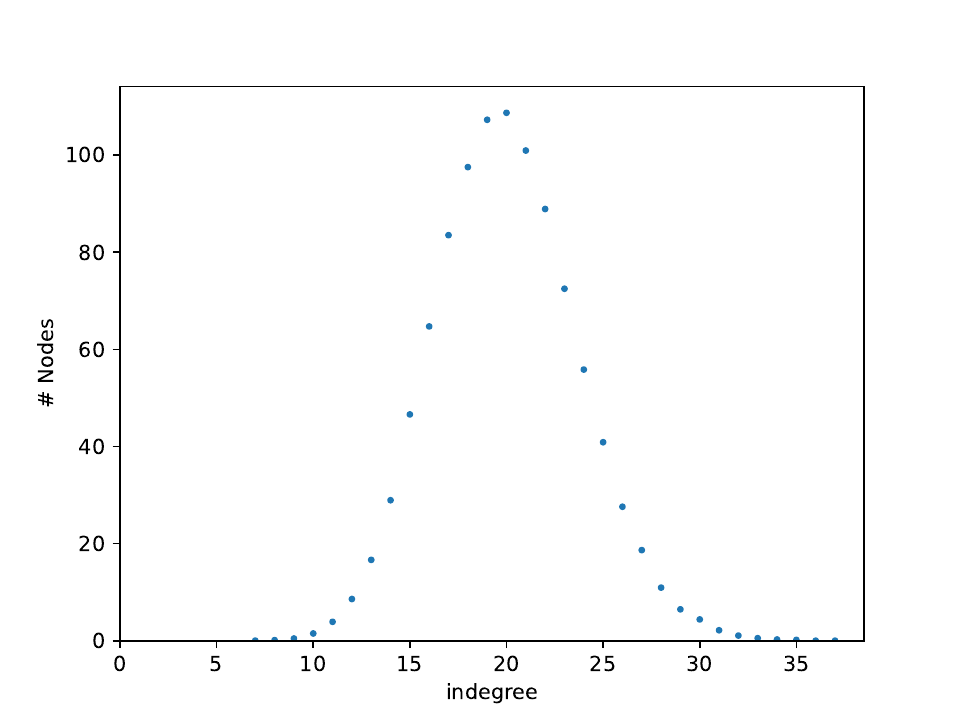}
    \caption{In-degree distribution of the network, after the run of the PROOFS algorithm for 1000 cycles.}
    \label{fig:ProofsDegreeDist}
\end{figure}

\begin{figure}[H]
    \includegraphics[width=0.85\columnwidth]{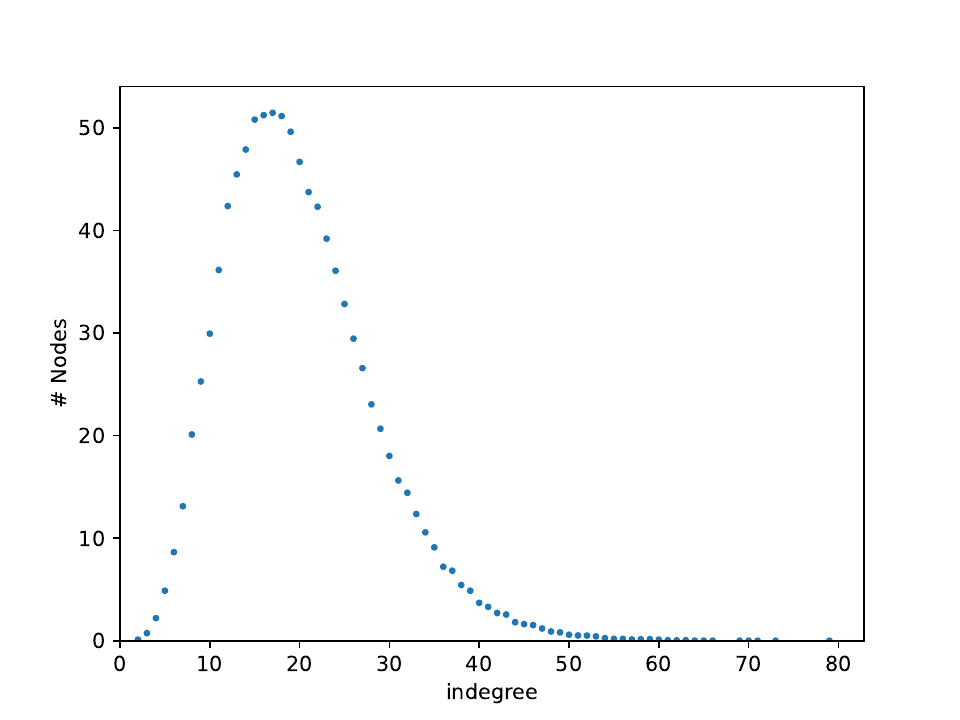}
    \caption{In-degree distribution of the network, after the run of the Newscast algorithm for 1000 cycles.}
    \label{fig:NewscastDegreeDist}
\end{figure}

\begin{figure}[H]
    \includegraphics[width=0.85\columnwidth]{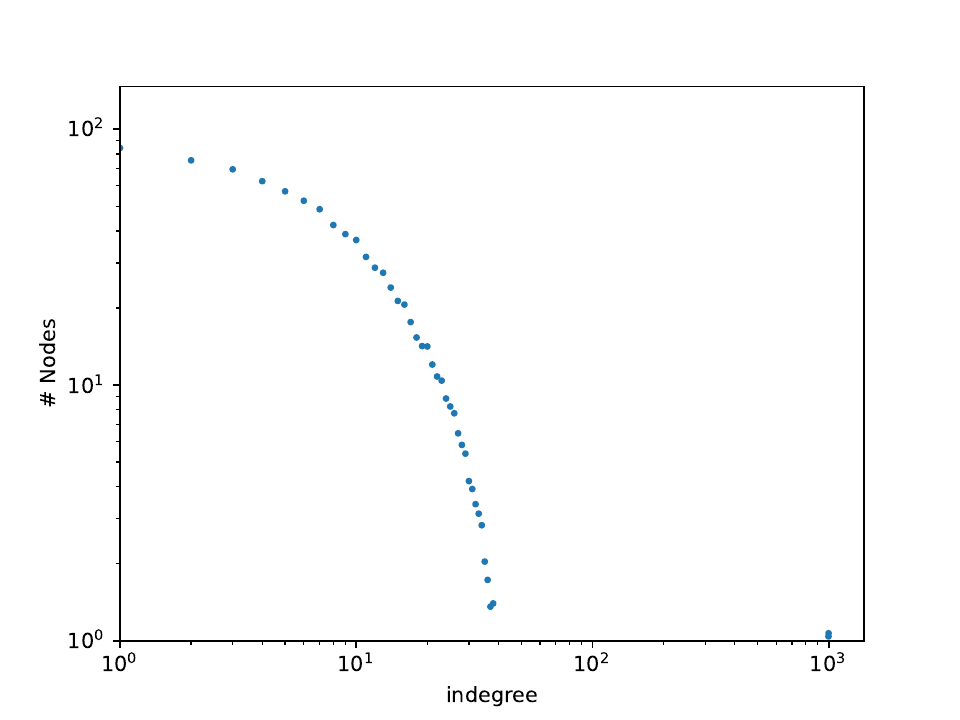}
    \caption{In-degree distribution of the network, after the run of the Phenix algorithm for 1000 cycles.}
    \label{fig:PhenixDegreeDist}
\end{figure}

\begin{figure}[H]
    \includegraphics[width=0.85\columnwidth]{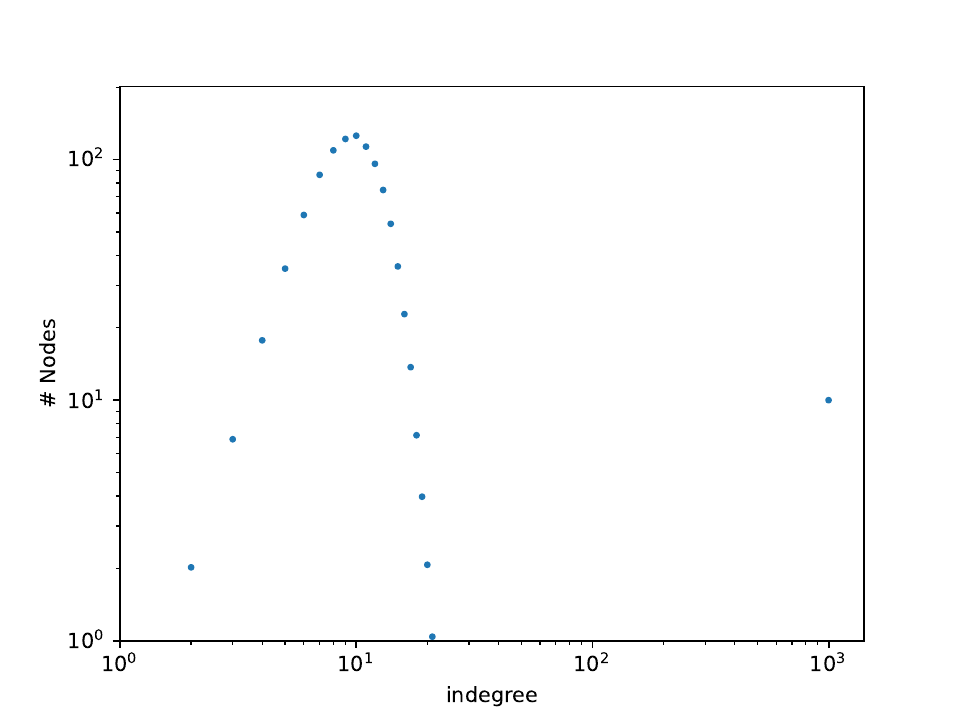}
    \caption{In-degree distribution of the network, after the run of the Elevator algorithm for 1000 cycles.}
    \label{fig:ElevatorDegreeDist}
\end{figure}

\begin{figure}[H]
    \includegraphics[width=0.85\columnwidth]{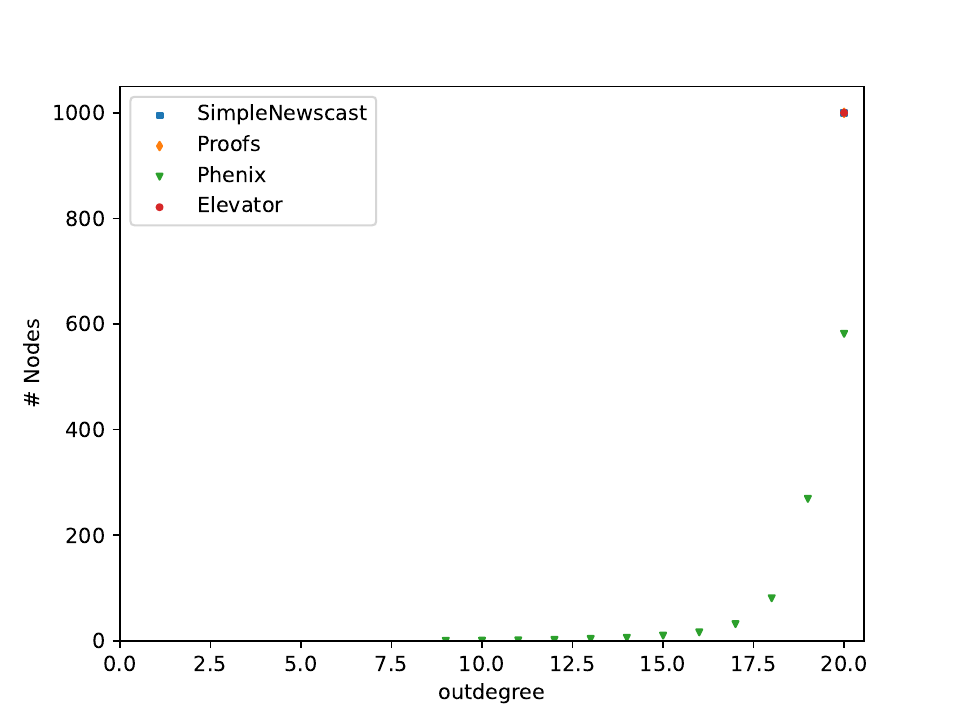}
    \caption{Out-degree distribution of the network, after the run of the algorithm for 1000 cycles, for each algorithm.}
    \label{fig:OutDegreeDist}
\end{figure}

\begin{figure}[H]
    \includegraphics[width=0.85\columnwidth]{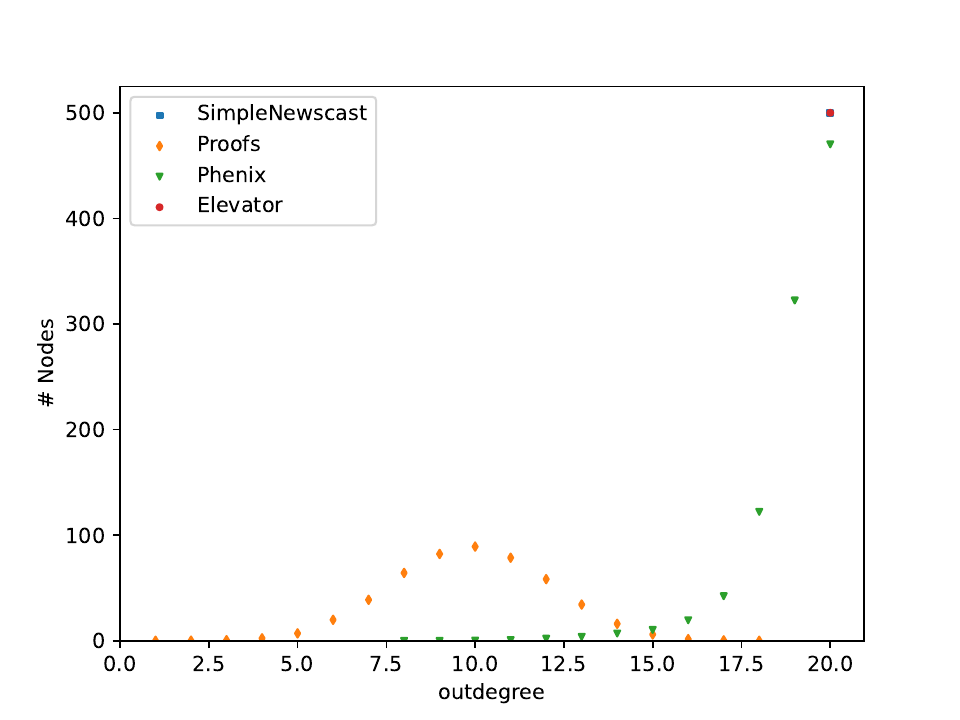}
    \caption{Out-degree distribution of the network, with a crash, for each algorithm.}
    \label{fig:OutDegreeDistCrash}
\end{figure}

\begin{figure}[H]
    \includegraphics[width=0.85\columnwidth]{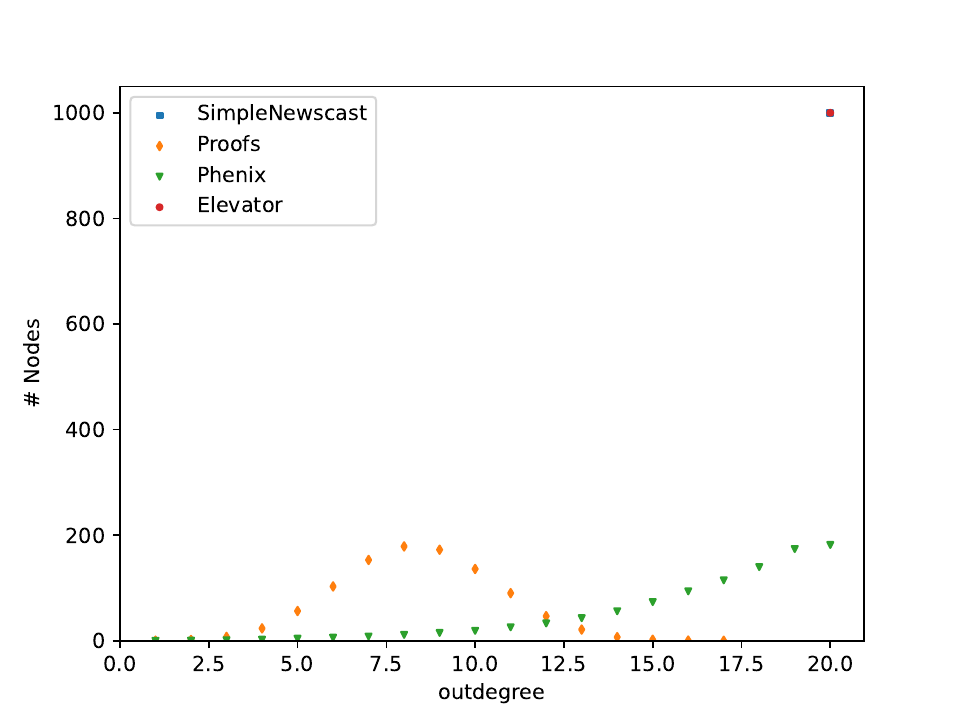}
    \caption{Out-degree distribution of the network, with churn, for each algorithm.}
    \label{fig:OutDegreeDistChurn}
\end{figure}

\begin{figure}[H]
    \includegraphics[width=0.85\columnwidth]{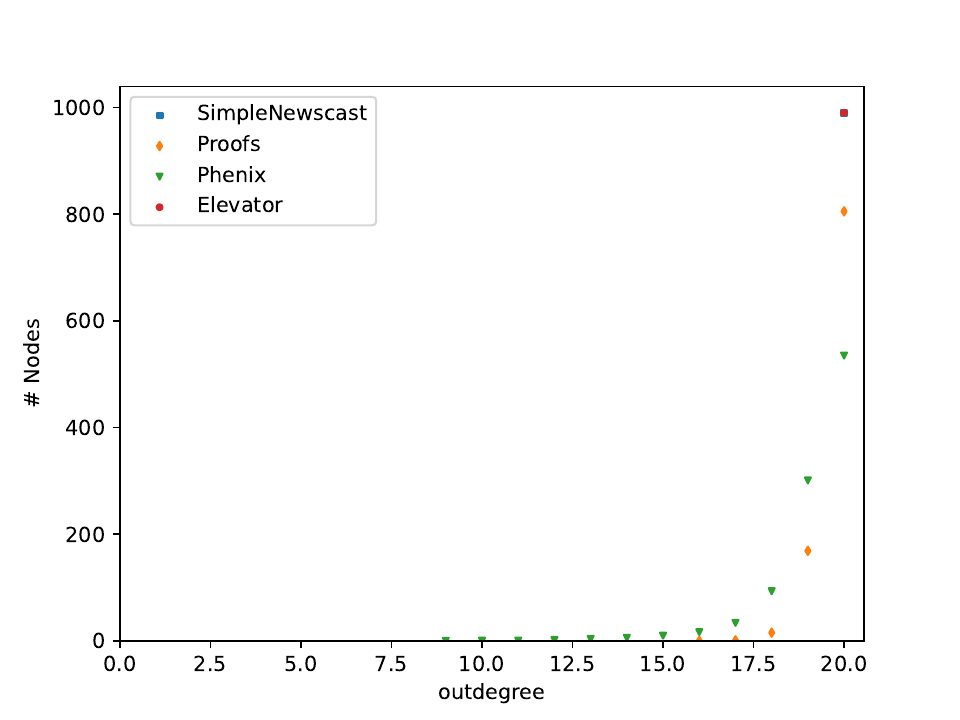}
    \caption{Out-degree distribution of the network, with a hub-targeted attack, for each algorithm.}
    \label{fig:OutDegreeDistCrashHub}
\end{figure}

\begin{figure}[H]
    \includegraphics[width=0.85\columnwidth]{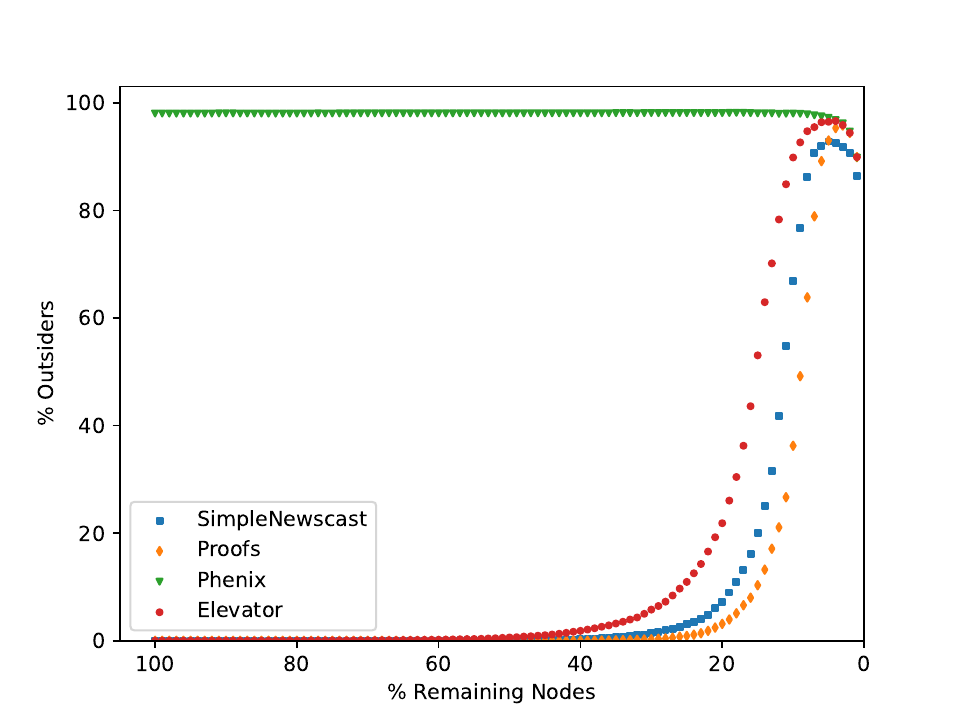}
    \caption{Analysis of the removal of all the nodes of the network one by one, and observing the number of nodes outside the biggest strongly connected cluster.}
    \label{fig:ComponentStrongFull}
\end{figure}

\begin{figure}[H]
    \includegraphics[width=0.85\columnwidth]{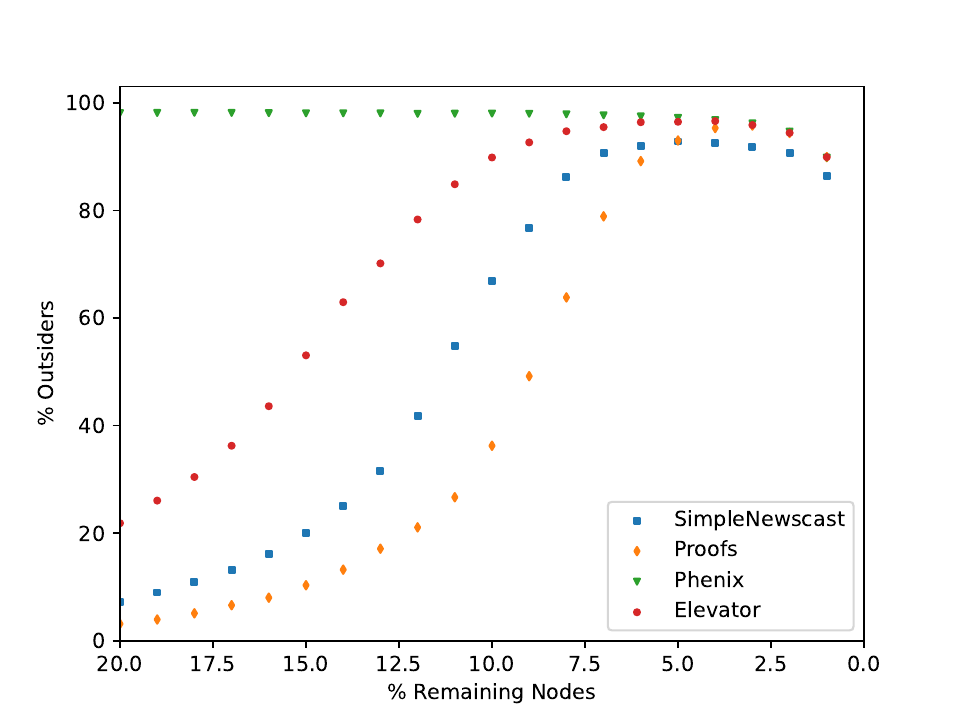}
    \caption{Analysis of the removal of all the nodes of the network one by one, and observing the number of nodes outside the biggest strongly connected cluster, zoom on the last 20\%.}
    \label{fig:ComponentStrongZoom}
\end{figure}

\begin{figure}[H]
    \includegraphics[width=0.85\columnwidth]{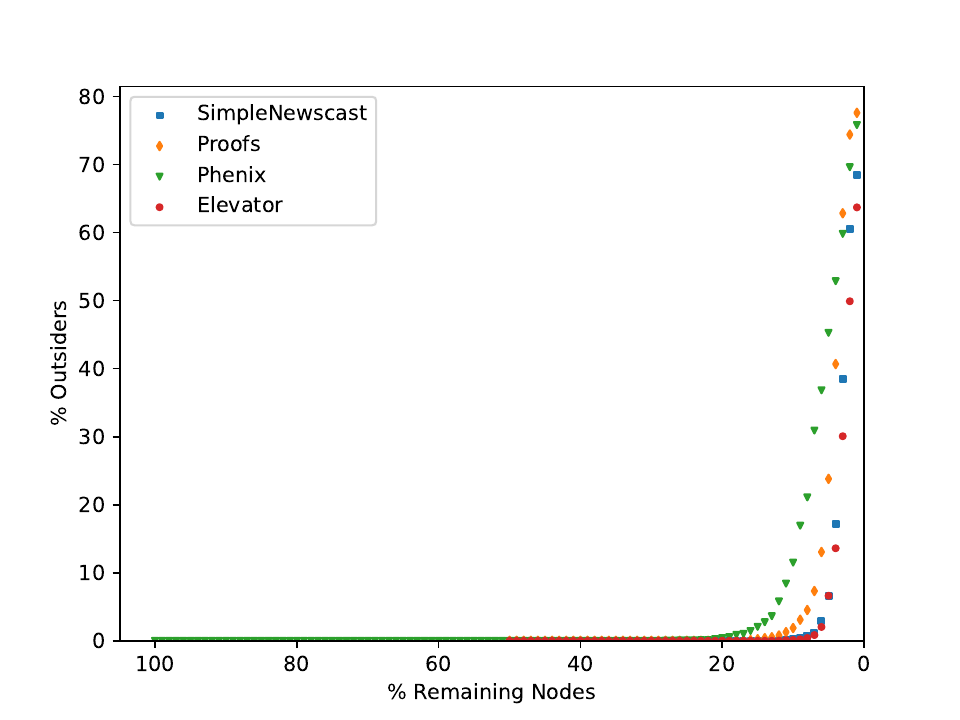}
    \caption{Analysis of the removal of all the nodes of the network one by one, and observing the number of nodes outside the biggest weakly connected cluster, with a 50\% crash.}
    \label{fig:ComponentFullCrash}
\end{figure}

\begin{figure}[H]
    \includegraphics[width=0.85\columnwidth]{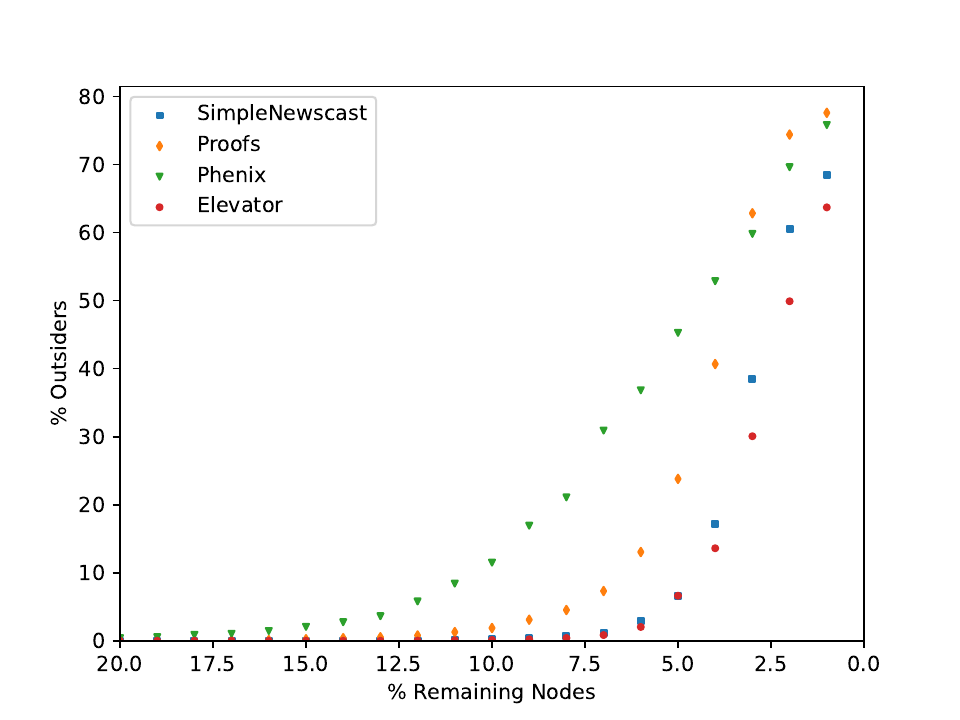}
    \caption{Analysis of the removal of all the nodes of the network one by one, and observing the number of nodes outside the biggest weakly connected cluster, with a 50\% crash, zoom on the last 20\%.}
    \label{fig:ComponentZoomCrash}
\end{figure}

\begin{figure}[H]
    \includegraphics[width=0.85\columnwidth]{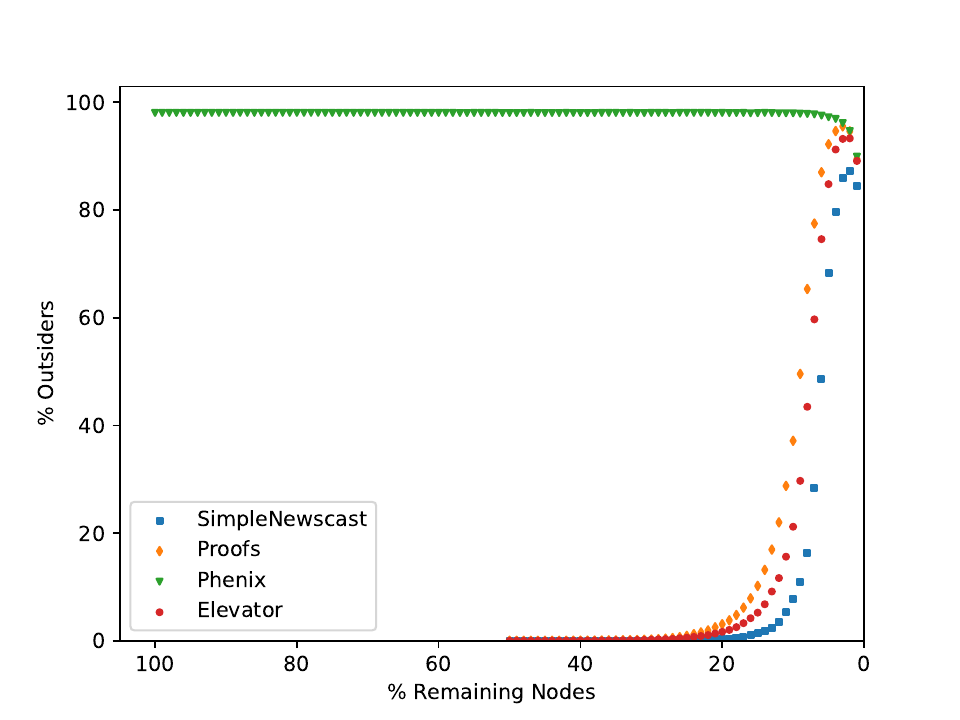}
    \caption{Analysis of the removal of all the nodes of the network one by one, and observing the number of nodes outside the biggest strongly connected cluster, with a 50\% crash.}
    \label{fig:ComponentStrongFullCrash}
\end{figure}

\begin{figure}[H]
    \includegraphics[width=0.85\columnwidth]{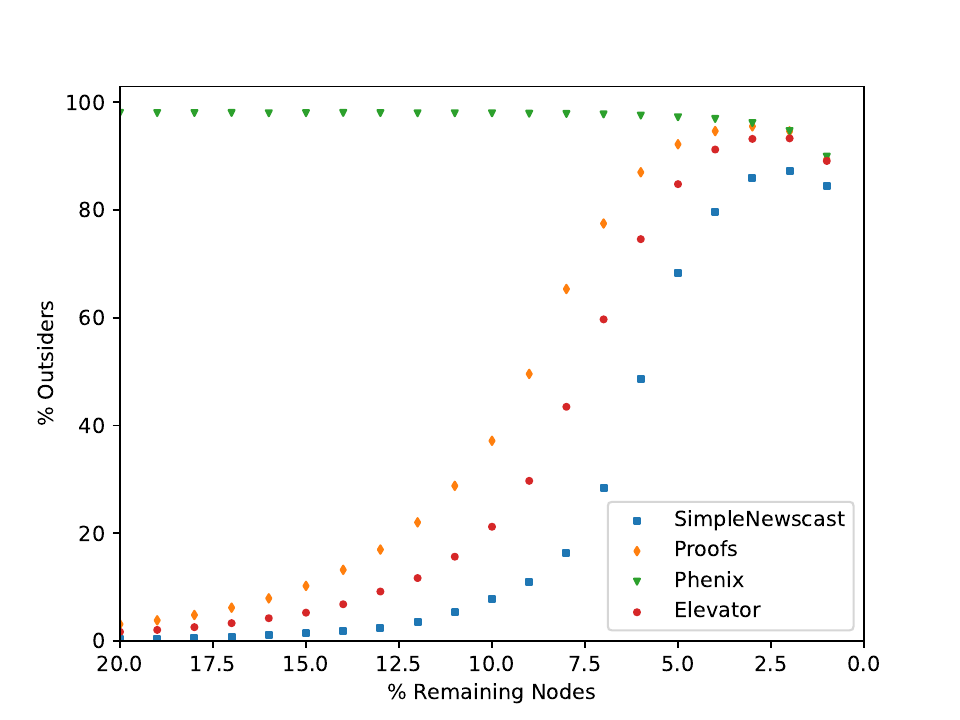}
    \caption{\small Analysis of the removal of all the nodes of the network one by one, and observing the number of nodes outside the biggest strongly connected cluster, with a 50\% crash, zoom on the last 20\%.}
    \label{fig:ComponentStrongZoomCrash}
\end{figure}

\begin{figure}[H]
    \includegraphics[width=0.85\columnwidth]{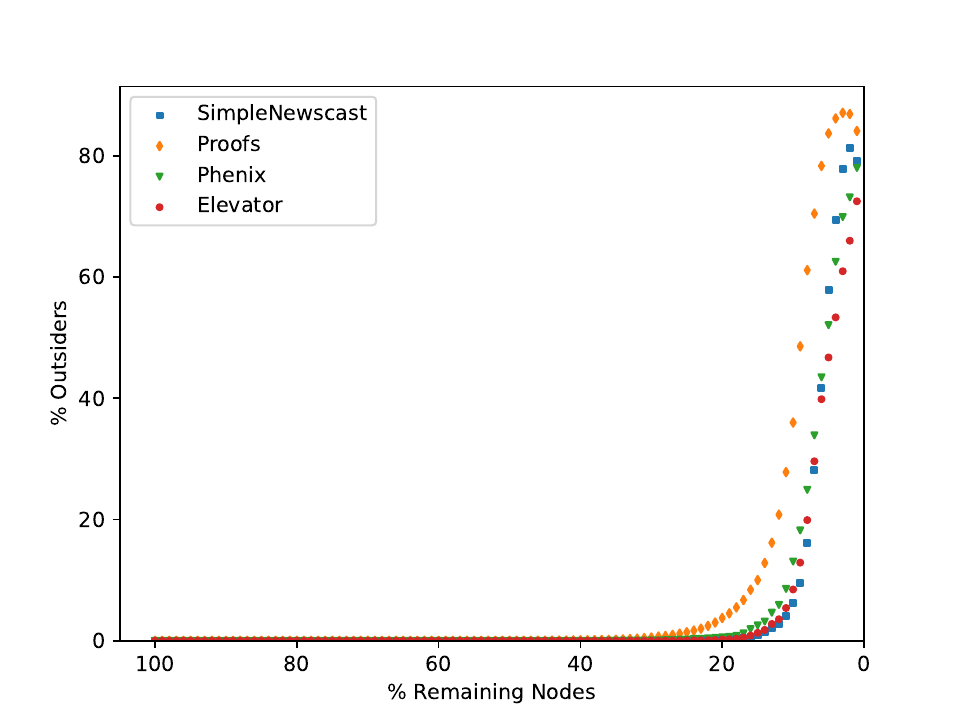}
    \caption{\small Analysis of the removal of all the nodes of the network one by one, and observing the number of nodes outside the biggest weakly connected cluster, with churn.}
    \label{fig:ComponentFullChurn}
\end{figure}

\begin{figure}[H]
    \includegraphics[width=0.85\columnwidth]{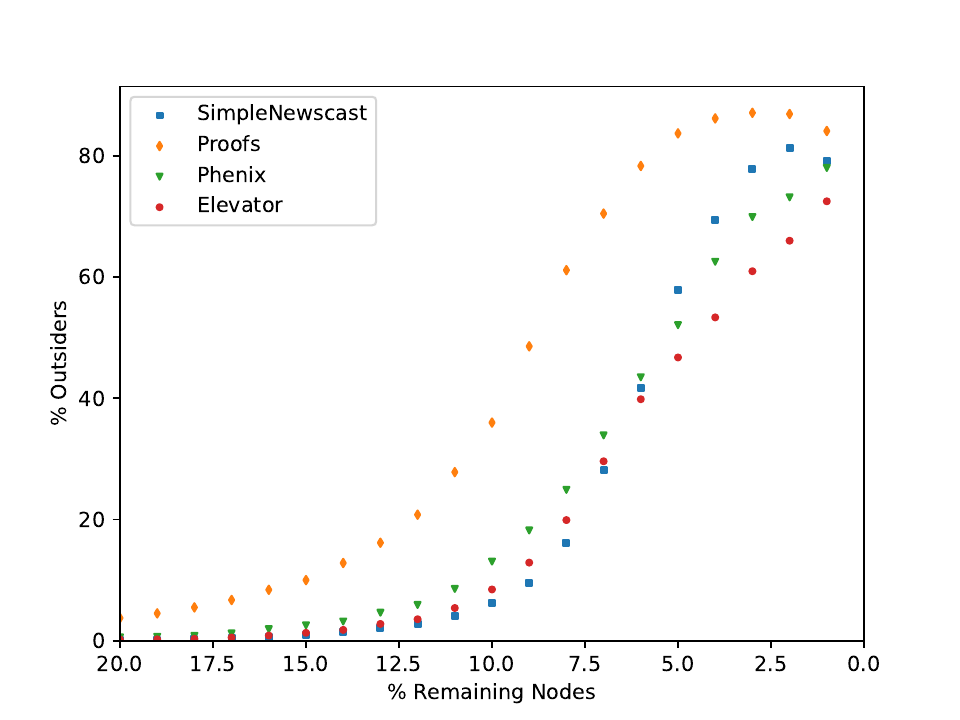}
    \caption{\small Analysis of the removal of all the nodes of the network one by one, and observing the number of nodes outside the biggest weakly connected cluster, with churn, zoom on the last 20\%.}
    \label{fig:ComponentZoomChurn}
\end{figure}

\begin{figure}[H]
    \includegraphics[width=0.85\columnwidth]{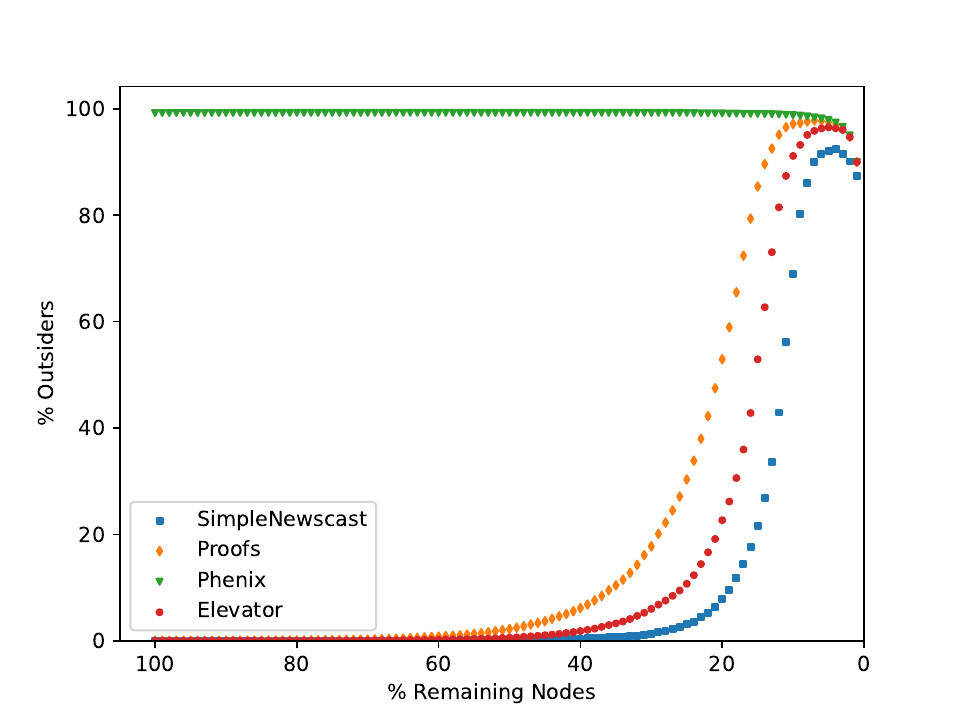}
    \caption{\small Analysis of the removal of all the nodes of the network one by one, and observing the number of nodes outside the biggest strongly connected cluster, with churn.}
    \label{fig:ComponentStrongFullChurn}
\end{figure}

\begin{figure}[H]
    \includegraphics[width=0.85\columnwidth]{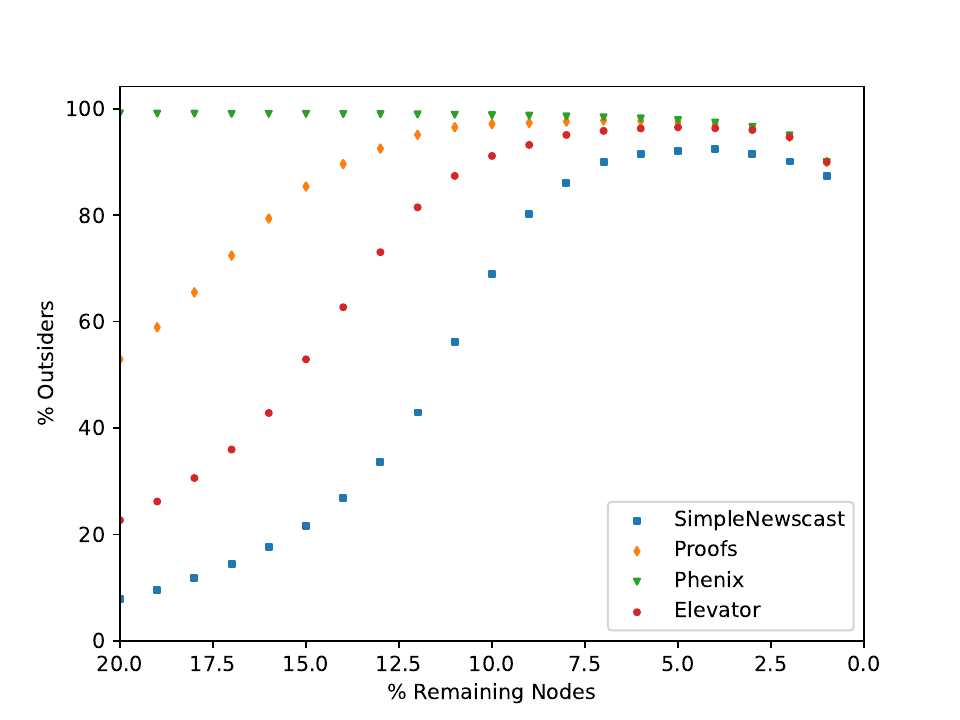}
    \caption{\small Analysis of the removal of all the nodes of the network one by one, and observing the number of nodes outside the biggest strongly connected cluster, with churn, zoom on the last 20\%.}
    \label{fig:ComponentStrongZoomChurn}
\end{figure}

\begin{figure}[H]
    \includegraphics[width=0.85\columnwidth]{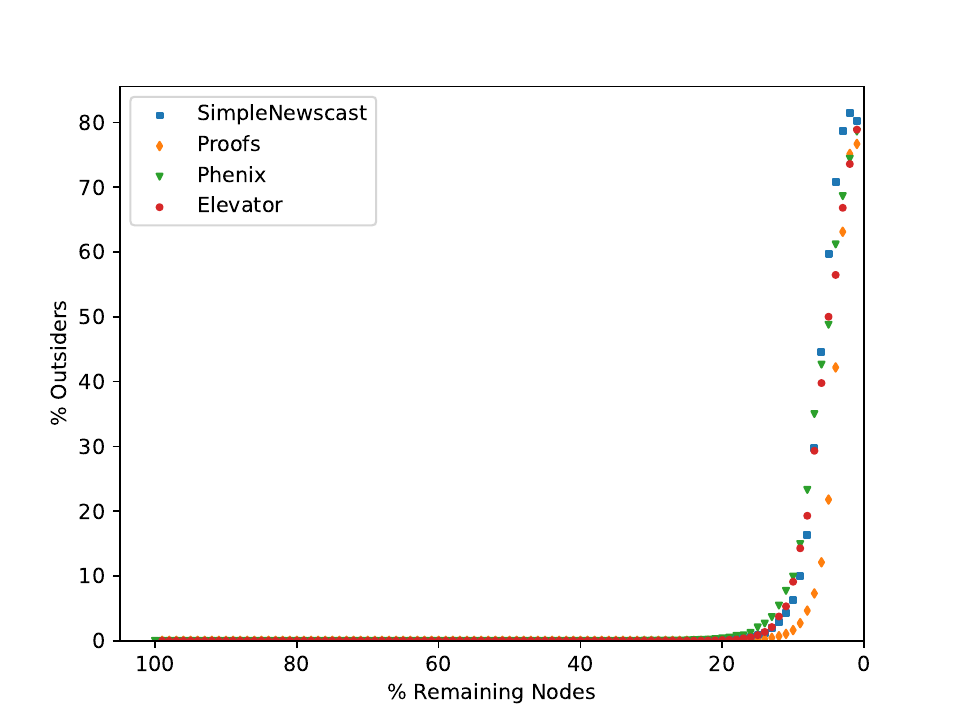}
    \caption{\small Analysis of the removal of all the nodes of the network one by one, and observing the number of nodes outside the biggest weakly connected cluster, with a hub-targeted attack.}
    \label{fig:ComponentFullCrashHub}
\end{figure}

\begin{figure}[H]
    \includegraphics[width=0.85\columnwidth]{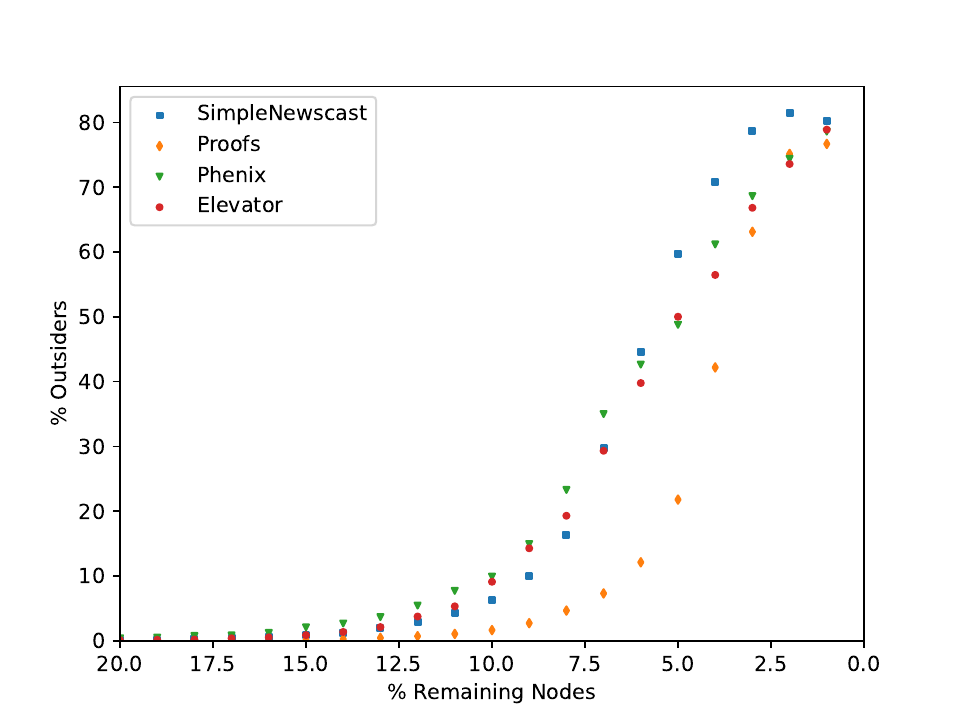}
    \caption{\small Analysis of the removal of all the nodes of the network one by one, and observing the number of nodes outside the biggest weakly connected cluster, with a hub-targeted attack, zoom on the last 20\%.}
    \label{fig:ComponentZoomCrashHub}
\end{figure}

\begin{figure}[H]
    \includegraphics[width=0.85\columnwidth]{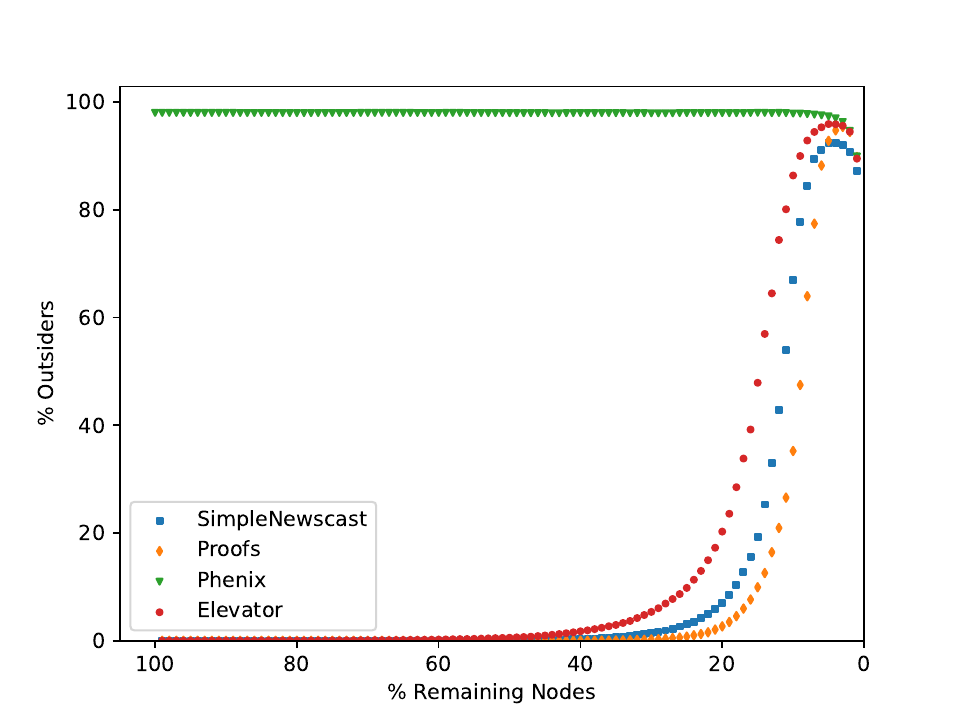}
    \caption{\small Analysis of the removal of all the nodes of the network one by one, and observing the number of nodes outside the biggest strongly connected cluster, with a hub-targeted attack.}
    \label{fig:ComponentStrongFullCrashHub}
\end{figure}

\begin{figure}[H]
    \includegraphics[width=0.85\columnwidth]{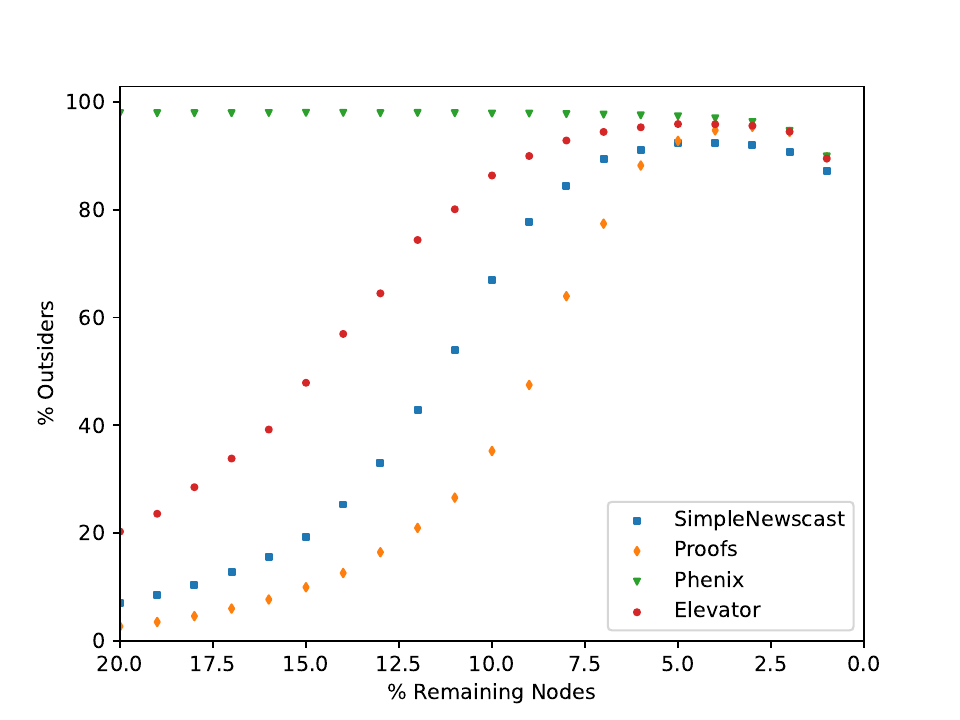}
    \caption{\small Analysis of the removal of all the nodes of the network one by one, and observing the number of nodes outside the biggest strongly connected cluster, with a hub-targeted attack, zoom on the last 20\%.}
    \label{fig:ComponentStrongZoomCrashHub}
\end{figure}

\end{document}